\documentclass[12pt]{article}
\usepackage{authblk}
\usepackage{graphicx}
\usepackage{amsmath,amssymb,mathabx}

\usepackage{lineno,hyperref}
\modulolinenumbers[5]

\usepackage{cleveref}
\usepackage{algorithm}
\usepackage{cite}
\graphicspath{./}

\newcommand{\f}[2]{\frac{#1}{#2}}
\newcommand{\la}{\langle}
\newcommand{\ra}{\rangle}
\newcommand{\lla}{\la\!\la}
\newcommand{\rra}{\ra\!\ra}
\newcommand{\de}{\partial}
\renewcommand{\mod}{{\rm mod}\,}
\newcommand{\B}{{\cal B}}

\title{A Geometric Monte Carlo Algorithm for the
    Antiferromagnetic Ising model with ``Topological'' Term at
    $\theta=\pi$}

\author[1]{V.~Azcoiti \thanks{azcoiti@azcoiti.unizar.es}}
\author[1,2]{G.~Cortese \thanks{cortese@unizar.es}}
\author[1]{E.~Follana \thanks{efollana@unizar.es}}
\author[3]{M.~Giordano \thanks{giordano@atomki.mta.hu}}

\affil[1]{Departamento de F\'isica Te\'orica, Universidad de
  Zaragoza, {\sl Calle Pedro Cerbuna 12, E-50009 Zaragoza, Spain.}}

\affil[2]{ Instituto de F\'isica Te\'orica, UAM/CSIC, {\sl Cantoblanco,
  E-28049 Madrid, Spain.}}

\affil[3]{Institute for Nuclear Research of the Hungarian
  Academy of Sciences (ATOMKI), {\sl Bem t\'er 18/c, H-4026 Debrecen,
    Hungary.}}

\begin{document}
\maketitle

\abstract{
In this work we study the two and three-dimensional antiferromagnetic
Ising model with an imaginary magnetic field $i\theta$ at
$\theta=\pi$. In order to perform numerical simulations of the system
we introduce a new geometric algorithm not affected by the sign
problem.  Our results for the $2D$ model are in agreement with the
analytical solution. We also present new results for the $3D$ model
which are qualitatively in agreement with mean-field predictions.
}

\section{Introduction}

Since its introduction many years ago, the Ising model has been a
prototype statistical system for studying phase transitions and
critical phenomena \cite{libroIsing}.  With the advent of the epoch of
computer numerical simulations to study statistical systems, this
model has become even more important as a test bench to develop new
algorithms. There are many interesting physical systems for which, due
to the sign problem, we do not have efficient numerical
algorithms. Some examples include QCD at finite density or with a
non-vanishing $\theta$ term. This situation has hindered progress in
such fields for a long time, and it is thus of great interest to study
novel simulation algorithms.  In the present work we develop and test
a new algorithm, which belongs to a class of ``geometric'' algorithms
\cite{worm1,endres1,endres2,worm2,wenger,worm31,worm32,worm33,worm34,worm35,worm36,worm37,worm0,worm41,worm42,worm43,worm44,worm45,worm46},
and which is applicable to the $D \ge 2$ dimensional antiferromagnetic
Ising model with an imaginary magnetic field $i\theta$ (see~\cite{Lee}
and~\cite{Matveev}) at $\theta=\pi$, with which we are able to solve
the sign problem that afflicts this model when using standard
algorithms. Preliminary results were presented in \cite{Pos}.

This paper is organized as follows. In Section \ref{model} we
introduce the Antiferromagnetic Ising model.  In Section
\ref{geometric} we derive our geometric representation of the
model. Section \ref{algorithm} is devoted to the construction of the
numerical algorithm. In Section \ref{numerical} we present the
numerical results and finally Section \ref{conclusions} contains our
conclusions. Some technical details on the ergodicity of the algorithm
as well as on the numerical analysis are contained in the appendices.

\section{The Antiferromagnetic Ising model with a topological term}
\label{model}

We consider the Ising model in $D\geq 2$ dimensions, defined on a
hypercubic lattice $\Lambda$ with an even number of sites $L = 2n$ in each
direction, and with either open or periodic boundary conditions.  The
Hamiltonian of the model is
\begin{equation}
  \label{eq:1}
  H[\{s_x\},J,B] = -J\sum_{(x,y)\in \B} s_x s_y -B\sum_{x} s_x\,.
\end{equation}
Here the spin variables are $s_x=\pm 1$, and the sum $\sum_{(x,y)\in
  \B}$ is over the pairs of sites $(x,y)$ that are nearest neighbors;
we denote the set of all such pairs by $\B$. Moreover, $J$ is the
coupling between nearest neighbors, and $B$ is an external magnetic
field.  The reduced Hamiltonian ${\cal H}=H/(k_{B}T)$, where $T$ is
the temperature and $k_{B}$ the Boltzmann constant, is written as
\begin{equation}
  \label{eq:2}
  {\cal H}[\{s_x\},F,h] = -F\sum_{(x,y)\in \B} s_x s_y -\f{h}{2}\sum_{x} s_x\,,
\end{equation}
with $F=J/(k_{B}T)$, $h=2B/(k_{B}T)$. As the total number of spins is
$L^D=(2n)^D$, and therefore even, the quantity $Q=\f{1}{2}\sum_{x}
s_x$ is an integer number, taking values between $-L^D/2$ and
$L^D/2$. $Q$ can then be thought of as playing the role of a
topological charge. It is then worth studying what happens for
imaginary values of the reduced magnetic field $h$, i.e., for
$h=i\theta$. The topological charge $Q$ is odd under the $Z_2$
transformation $s_x\to -s_x$ $\forall x$: while at $\theta=0$ the
system is symmetric under this transformation, for $\theta\neq 0,\pi$
the $Z_2$ symmetry of the system is explicitly broken. At $\theta=\pi$
the contribution of the topological charge to the Boltzmann factor
amounts to
\begin{equation}
  \label{eq:3}
  e^{i\pi Q} = (-1)^Q = (-1)^{-Q}\,,
\end{equation}
i.e., this contribution is $Z_2$ invariant, and therefore the $Z_2$
symmetry is restored; it has to be checked if it is spontaneously
broken or not.

\section{Geometric representation of the model}
\label{geometric}

\subsection{Partition function}

We will now introduce a geometric representation for the model at
\mbox{$h = i \theta = i\pi$.} Let us rewrite the partition function of
the system,
\begin{equation}
  \begin{aligned}
  Z(F,\theta=\pi) &=   \sum_{s_x=\pm 1} e^{F\sum_{(x,y)\in \B} s_x
    s_y + i\f{\pi}{2}\sum_{z} s_z}    \\
&= \sum_{s_x=\pm 1} \prod_{(x,y)\in \B} e^{F s_x s_y} \prod_z
i^{s_z} =  \sum_{s_x=\pm 1} \prod_{(x,y)\in \B} e^{F s_x s_y} \prod_z
i{s_z}  \\
&= i^{V}\sum_{s_x=\pm 1} \prod_{(x,y)\in \B} [\cosh(F s_x s_y) +
\sinh(F s_x s_y)] \prod_z {s_z} \\
&= \sum_{s_x=\pm 1} \prod_{(x,y)\in \B} [\cosh(F) +
\sinh(F)s_x s_y] \prod_z {s_z}
\,,
  \end{aligned}
  \label{partition1}
\end{equation}
where we have taken into account that the volume of the system, $V =
L^D$, is a multiple of $4$, and that $\cosh(-x)= \cosh(x)$,
$\sinh(-x)= - \sinh(x)$.  Note now the following symmetry of this
partition function. Define the two staggered lattices
$\Lambda^{(1,2)}$ as follows:
\begin{equation}
  \label{eq:5}
  \begin{aligned}
  \Lambda^{(1)}&=\{x=(i_1, \ldots, i_D)\in\Lambda ~|~
  (i_1+\cdots+i_D)\mod 2=0\} \,,\\ \Lambda^{(2)}&=\{x=(i_1, \ldots, i_D)\in\Lambda
  ~|~ (i_1+\cdots+i_D)\mod 2=1\} \,.
  \end{aligned}
\end{equation}
Nearest neighbors always belong to different staggered lattices; as a
consequence, if we change variables in the sum in
Eq.~\eqref{partition1} by changing the sign of all the spins in only
one of the two staggered lattices, say, $s_x\to -s_x\,\forall
x\in\Lambda^{(2)}$, Eq.~\eqref{partition1} becomes
\begin{equation}
  \label{eq:6}
  \begin{aligned}
  Z(F,\theta=\pi) &= (-1)^{\f{L^D}{2}}\sum_{s_x=\pm 1} \prod_{(x,y)\in \B} [\cosh(F) -
\sinh(F)s_x s_y] \prod_z {s_z}\\
&= \sum_{s_x=\pm 1} \prod_{(x,y)\in \B} [\cosh(F) -
\sinh(F)s_x s_y] \prod_z {s_z}\,.
  \end{aligned}
\end{equation}
Therefore, $Z(F,\theta=\pi)=Z(-F,\theta=\pi)$, i.e., at $\theta=\pi$
the ferromagnetic and antiferromagnetic models are essentially equivalent. In
conclusion, we can write
\begin{equation}
  \label{eq:7}
  Z(F,\theta=\pi)= \sum_{s_x=\pm 1} \prod_{(x,y)\in \B}
  [\cosh(|F|) + 
  \sinh(|F|)s_x s_y] \prod_z {s_z}\,.
\end{equation}
Let us denote by $\B!$ the power set of $\B$, i.e.,
\begin{equation}
  \label{eq:7ter}
  \B! = \{ b~|~b\subseteq \B\}\,.
\end{equation}
A subset $b$ can be seen as a configuration of ``active bonds'' (we
will sometimes also refer to active bonds as ``dimers'') between
neighboring sites. Let ${\cal N}[b]$ be the number of elements of $b$,
i.e., the number of active bonds; clearly, $\bar{\cal N}[b]={\cal
  N}[\B]-{\cal N}[b]$ is the number of inactive bonds.  Finally, we
define the quantity
\begin{equation}
  \label{eq:7quater}
  \pi_x[(y,z)]= \left\{
    \begin{aligned}
      &1 &&& &\text{if}~x=y~\text{or}~x=z\,,\\
      &0 &&& &\text{otherwise}\,,
    \end{aligned}\right.
\end{equation}
and let $\pi_x[b]$ be the number of bonds in $b$ that ``touch'' $x$,
\begin{equation}
  \label{eq:7quinquies}
  \pi_x[b] = \sum_{(y,z)\in b}\pi_x[(y,z)]\, .
\end{equation}
Armed with this notation, we can rewrite the product over pairs of
neighboring sites in Eq.~\eqref{eq:7} as follows,
\begin{multline}
  \label{eq:7sexties}
  \prod_{(x,y)\in \B} [\cosh(|F|) + \sinh(|F|)s_x s_y] =  \\  
  \sum_{b\in \B!} \cosh(|F|)^{\bar{\cal N}[b]} \sinh(|F|)^{{\cal
      N}[b]}\prod_{(x,y)\in b} s_x s_y\,. 
\end{multline}
The sum over spin configurations in $Z$ vanishes unless each spin
appears an even number of times in the product above, and gives a
factor of 2 per spin otherwise, that is,
\begin{equation}
  \label{eq:7septies}
\sum_{s_x=\pm 1}  \prod_{(x,y)\in b} s_x s_y \prod_z
s_z = \left\{
  \begin{aligned}
    &0 &&& &\text{if}~\exists x \owns \pi_x[b]\,\mod 2 = 0\,,\\
    &2^{L^D} &&& &\text{otherwise}\,.
  \end{aligned}\right\}
\end{equation}
Summarizing,
\begin{equation}
  \label{eq:7opties}
  \begin{aligned}
    Z(F,\theta=\pi)&=  2^{L^D}\sum_{\substack{b\in \B!,\\
        \{\pi_x[b]\mod 2 = 1 \forall x\}}} \cosh(|F|)^{\bar{\cal N}[b]}
    \sinh(|F|)^{{\cal N}[b]} \\ 
    &= 2^{L^D}\cosh(|F|)^{{\cal N}[\B]}\sum_{\substack{b\in
        \B!,\\ 
        \{\pi_x[b]\mod 2 = 1 \forall x\}}}  \tanh(|F|)^{{\cal N}[b]}\,.
  \end{aligned}
\end{equation}
This is the geometric representation that we will use in our
algorithm.

\subsection{Observables}

It is useful to generalize the partition function to the case of
variable couplings, i.e,
\begin{equation}
  \label{eq:8}
  {\cal H}\left[\{s_x\},\{F_{xy}\}, \theta = \pi \right] = 
-\sum_{(x,y)\in \B}F_{xy}\, s_x s_y
  -i\f{\pi}{2}\sum_{x} s_x\,. 
\end{equation}
This allows us to calculate all the correlation functions for an even
number of spins $\la s_{x_1} s_{x_2}\ldots s_{x_{2k}}
\ra$,\footnote{Correlation functions with an odd number of spins are
  automatically zero.} by taking derivatives with respect to $F_{xy}$
for an appropriate set of $(x,y)$. Indeed, choosing a set of paths
${\cal C}_1,\ldots {\cal C}_k$ connecting the spins pairwise (there
are no restrictions on these paths), and then performing derivatives
with respect to all the pairs $(x,y)$ appearing in those paths (if a
pair appears $m(x,y)$ times, one has to take the $m(x,y)$-th
derivative with respect to the corresponding coupling),
\begin{multline}
  \label{eq:9bis}
 \la s_{x_1} s_{x_2}\ldots s_{x_{2k}} \ra = Z^{-1}(\{F_{xy} = F\},\theta=\pi)
\\ \times \left\{ \left[\prod_{(x,y)\in \cup_j {\cal C}_j} \f{\de^{m(x,y)}}{\de
     F_{xy}^{m(x,y)}}\right]Z(\{F_{xy}\},\theta=\pi)\right\}\Bigg|_{\{F_{xy}\}=\{F\}}
 \,.
\end{multline}
The geometric representation for the partition function with variable
couplings is similar to the one obtained for constant coupling, and
the final result takes the form
\begin{equation}
  \label{eq:11}
    Z(\{F_{xy}\},\theta=\pi)=   2^{L^D}\sum_{\substack{b\in \B!,\\
       \{ \pi_x[b]\mod 2 = 1 \forall x\}}} 
    \prod_{\substack{(x,y) \\
      (x,y) \notin b}}\cosh(F_{xy})  \prod_{\substack{(x,y) \\
      (x,y) \in b}} \sinh(F_{xy})\,.
\end{equation}
Taking derivatives with respect to $F_{x_1 y_1},\ldots, F_{x_l y_l}$,
and carrying them inside the summation over $b$, one obtains an extra
factor $\coth( F_{x_i y_i})$ if $(x_i, y_i)$ is an active bond of the
configuration, i.e., if $(x_i, y_i)\in b$, or $\tanh(F_{x_i,y_i})$ if
$(x_i,y_i)$ is inactive, i.e., $(x_i, y_i)\notin b$:
\begin{equation}
  \label{eq:12}
  \begin{aligned}
&\left[\prod_{i=1}^l \f{\de}{\de F_{x_i
      y_i}}\right]Z(\{F_{xy}\},\theta=\pi) = \\ &
    2^{L^D}\sum_{\substack{b\in \B !,\\ 
        \{\pi_x[b]\mod 2 = 1 \forall x\}}} 
   \prod_{i=1}^l\left[ \tanh( F_{x_i y_i})(1-\delta_b(x_i, y_i))
    + \coth (F_{x_i y_i}) \delta_b(x_i, y_i) \right]\\ & \times
  \,\,\,\,\,\,\,\,\,  \prod_{\substack{(x,y) \\
      (x,y) \notin b}}\cosh(F_{xy})  \prod_{\substack{(x,y)\\
      (x,y) \in b}} \sinh(F_{xy})\,,
  \end{aligned}
\end{equation}
where $\delta_b( x, y)=1$ if $(x,y)\in b$, and 0 otherwise. Noting
that
\begin{multline}
  \label{eq:12bis}
  \tanh( F_{x_i y_i})(1-\delta_b(x_i, y_i))
    + \coth (F_{x_i y_i}) \delta_b(x_i, y_i) \\ = \tanh( F_{x_i
      y_i})^{1-2\delta_b(x_i, y_i)}\,,
\end{multline}
and setting $F_{xy}=F$, $\forall (x,y)$, we finally obtain
\begin{equation}
  \label{eq:13}
  \begin{aligned}
&\left[\prod_{i=1}^l \f{\de}{\de F_{x_i
      y_i}}\right]Z(\{F_{xy}\},\theta=\pi)\bigg|_{\{F_{xy}\}=\{F\}}= \\ & 
2^{L^D}\sum_{\substack{b\in \B!,\\ 
       \{\pi_x[b]\mod 2 = 1 \forall x\}}} \prod_{i=1}^l\left[ \tanh(
      F)^{1-2\delta_b(x_i, y_i)} \right] 
    \prod_{\substack{(x,y) \\
      (x,y) \notin b}}\cosh(F)  \prod_{\substack{(x,y)\\
      (x,y) \in b}} \sinh(F) \\
&=2^{L^D} \cosh(F)^{{\cal N}[\B]}\sum_{\substack{b\in \B!,\\ 
        \{\pi_x[b]\mod 2 = 1 \forall x\}}} \prod_{i=1}^l\left[ \tanh(
      F)^{1-2\delta_b(x_i, y_i)} \right]\tanh(F)^{{\cal N}[b]}\,. 
  \end{aligned}
\end{equation}
Finally, denoting by $\Delta[b;\{x_i,y_i\}]=\sum_{i=1}^l
\delta_b(x_i,y_i)$, and dividing by the partition function, we obtain
\begin{equation}
  \label{eq:14}
  \begin{aligned}
    &  \la s_{x_1} s_{x_2}\ldots s_{x_{2k}} \ra =  
\f{\displaystyle\sum_{\substack{b\in \B!,\\ 
        \{\pi_x[b]\mod 2 = 1 \forall x\}}}
    \tanh(F)^{l-2\Delta[b;\{x_i,y_i\}]}\tanh(F)^{{\cal N}[b]} }
{\displaystyle\sum_{\substack{b\in \B!,\\ 
       \{\pi_x[b]\mod 2 = 1 \forall x\}}}
    \tanh(F)^{{\cal N}[b]} } = \\
& \,\,\,\,\,\, \lla
  \tanh(F)^{l-2\Delta[b;\{x_i,y_i\}]} \rra\,,
  \end{aligned}
\end{equation}
where the symbol $\lla \ldots \rra$ indicates the average taken with
the probability distribution\footnote{The proof of ergodicity in
  \ref{ergodicity} shows that ${\cal N}[b]$ is even for any admissible
  configuration.}
\begin{equation}
P(b)=
\frac{\tanh(F)^{{\cal N}[b]}}{\displaystyle\sum_{\substack{b\in \B!,\\ 
       \{\pi_x[b]\mod 2 = 1 \forall x\}}}
    \tanh(F)^{{\cal N}[b]} }\,.
 \end{equation}   
As an example, let us write down the two-point correlation function
$\la s_x s_y \ra$ for $x$ and $y$ lying on the same lattice axis,
e.g., $y=x+l \hat 1$. Let ${\cal C}$ be a path connecting $x$ and $y$
on the lattice; the simplest choice is a straight-line path ${\cal C}=
\cup_{i=1}^{l} (x_i,y_i)$ with $y_i=x_{i+1}$, $x_i=x+(i-1)\hat 1$
($i=1,l+1$), $x_1=x$ and $y_l=y$. We have
\begin{equation}
  \label{eq:14bis}
  C(l,F) \equiv \la s_x s_{x+l\hat 1} \ra  =  \lla
  \tanh(F)^{l-2\Delta[b;\{x_i,y_i\}]} \rra\,,
\end{equation}
where $\Delta[b;\{x_i,y_i\}]$ is the number of active bonds in
configuration $b$ along the straight-line path connecting $x$ and $x+l\hat
1$. However, we stress the fact that the specific choice of the path
is irrelevant, as they are all equivalent, as long as the endpoints
are fixed.

We can also obtain expressions for bulk observables, such as the
energy density and the specific heat, by taking derivatives of the
partition function with respect to the coupling constant. Using the
geometric representation it is easy to see that such observables are
directly related to the average number of active bonds and its
fluctuations. We denote by $B={\cal N}[b]$ the total number of active
bonds, $H_0$ is defined as
\begin{equation}
H_{0}=-J\sum_{(x,y) \in {\cal B}} s_{x}s_{y},
\end{equation} 
and we denote the fluctuations in these quantities by $\Delta H_{0}=
H_{0} - \langle H_{0} \rangle$, $\Delta B=B-\lla B \rra$. Then
we have the following relations for the energy density $\varepsilon$
and the specific heat $c_{V}$ (we only show for simplicity the
relation in the case of periodic boundary conditions):
\begin{eqnarray}
\label{equation:energy}
\varepsilon \equiv -\left\langle \frac{H_{0}}{k_{B}TV}\right\rangle = DF\tanh F +
\frac{2F}{\sinh 2F}\left\langle\!\!\left\langle \frac{B}{V} \right\rangle\!\!\right\rangle
\\
\begin{aligned}
c_{V} \equiv \left\langle \frac{1}{V}\left ( \frac{\Delta H_{0}}{k_{B}T} \right )^{2} \right\rangle =  \frac{DF^{2}}{(\cosh F)^{2}} - \left
  (\frac{2F}{\sinh 2F} \right)^{2} \cosh 2F \left\langle\!\!\left\langle \frac{B}{V}
 \right\rangle\!\!\right\rangle   + \\ \left(\frac{2F}{\sinh 2F} \right)^{2}
\left\langle\!\!\left\langle \frac{1}{V}\left (\Delta B\right )^{2} \right\rangle\!\!\right\rangle\,,
\label{eq:endens_specheat}
\end{aligned}
\label{equation:specific}
\end{eqnarray}
where $D$ is the dimensionality of the system.  For small values of the
coupling $F$ the energy density and the specific heat are
approximately equal to the average occupation number of the dimers and to its
fluctuations:
\begin{eqnarray}
\varepsilon \simeq \left\langle\!\!\left\langle \frac{B}{V} \right\rangle\!\!\right\rangle \\ 
c_{V} \simeq \left\langle\!\!\left\langle \frac{1}{V}\left (\Delta B\right )^{2} \right\rangle\!\!\right\rangle -
\left\langle\!\!\left\langle \frac{B}{V} \right\rangle\!\!\right\rangle
\end{eqnarray}

\section{Monte Carlo algorithm}
\label{algorithm}

In order to perform calculations by means of Monte Carlo methods, we
need an efficient algorithm to explore the space of configurations,
that as we have seen is given, in the geometric representation, by
\begin{equation}
  \label{eq:s1}
  \tilde \B \equiv \{b\in \B!~|~
        \pi_x[b]\,\mod 2 = 1 \,\forall x\}\,,
\end{equation}
or, in words, by those configurations for which the number of active
bonds touching any site $x$ of the lattice is odd. We will call the
set $\tilde{\B}$ the set of {\it admissible configurations}. For
simplicity we will describe in detail the algorithm only for $D=2$, as it can be easily generalized to $D > 2$.

As far as numerical simulations are concerned, it is enough if we know
at least one admissible configuration, and a set of updating rules
that take us from one admissible configuration to another and that
satisfy detailed balance and ergodicity.

\begin{figure}[t]
  \centering
  \begin{picture}(80,80)(-40,-15)
  \put(0,0){\circle*{6}}
  \put(40,0){\circle*{6}}
  \put(40,40){\circle*{6}}
  \put(0,40){\circle*{6}}
  \put(0,0){\line(1,0){10}}
  \put(15,0){\line(1,0){10}}
  \put(30,0){\line(1,0){10}}
  \put(40,0){\line(0,1){10}}
  \put(40,15){\line(0,1){10}}
  \put(40,30){\line(0,1){10}}
  \put(0,40){\line(1,0){10}}
  \put(15,40){\line(1,0){10}}
  \put(30,40){\line(1,0){10}}
  \put(0,0){\line(0,1){10}}
  \put(0,15){\line(0,1){10}}
  \put(0,30){\line(0,1){10}}
  \put(18,-15){$1$} 
  \put(48,16){$2$} 
  \put(18,50){$3$}
  \put(-12,16){$4$}  
  \put(18,18){$x^*$} 
\end{picture}
  \caption{Site $x^*$ on the dual lattice in $2D$. The dashed lines indicate
    bonds whose state is not specified.}
  \label{fig:1}
\end{figure}
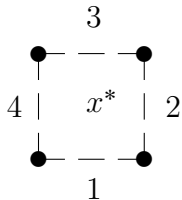

To describe the updating rules we found convenient to work with the
dual lattice (equivalently, the set of squares of the original
lattice). Let us denote a point in the dual lattice by $x^*$ , and let
us enumerate the four links of the corresponding square in
anticlockwise order, as shown in Fig.~(1). To describe a configuration
of bonds, we introduce for each site of the dual lattice a vector with
four components, $A(x^*)\equiv (A_1(x^*), A_2(x^*), A_3(x^*),
A_4(x^*))$, defined such that $A_i(x^*)=1$ if the corresponding bond
is active and $A_i(x^*)=0$ if it is inactive. The bond configuration
is completely specified by the dual lattice field $A(x^*)$, although
this description is redundant: indeed, one has that
$A_3(x^*)=A_1(x^*+\hat 1)$ and $A_2(x^*)=A_4(x^*+\hat 2)$. In Table 1
we show all the possible configurations of a given square on the
lattice, i.e., a point $x^*$ in the dual lattice. The graphs $S(x^*)$
corresponding to these configurations, and the number of active bonds,
$w(x^*)=\sum_{i=1}^4 A_i(x^*)$, are also shown.

\begin{table}[h!]
  \centering
  \begin{tabular}{c|c|c|c|c|c}
 $S(x^*)$ & $w(x^*)$ & $A(x^*)$ & $S(x^*)$ & $w(x^*)$ & $A(x^*)$ 
 \\ \hline
\begin{picture}(20,20)(0,-2.5)
  \put(0,0){\circle*{3}}
  \put(10,0){\circle*{3}}
  \put(10,10){\circle*{3}}
  \put(0,10){\circle*{3}}
\end{picture}
& 0 & 
$(0,0,0,0)$ &
\begin{picture}(20,20)(0,-2.5)
  \put(0,0){\circle*{3}}
  \put(10,0){\circle*{3}}
  \put(10,10){\circle*{3}}
  \put(0,10){\circle*{3}}
  \put(10,0){\line(0,1){10}}
  \put(10,10){\line(-1,0){10}}
\end{picture}
& 2  & 
$(0,1,1,0)$ \\ \hline
\begin{picture}(20,20)(0,-2.5)
  \put(0,0){\circle*{3}}
  \put(10,0){\circle*{3}}
  \put(10,10){\circle*{3}}
  \put(0,10){\circle*{3}}
  \put(0,0){\line(1,0){10}}
\end{picture}
& 1  & $(1,0,0,0)$ &
\begin{picture}(20,20)(0,-2.5)
  \put(0,0){\circle*{3}}
  \put(10,0){\circle*{3}}
  \put(10,10){\circle*{3}}
  \put(0,10){\circle*{3}}
  \put(10,0){\line(0,1){10}}
  \put(0,10){\line(0,-1){10}}
\end{picture}
& 2  & 
$(0,1,0,1)$ \\ \hline
\begin{picture}(20,20)(0,-2.5)
  \put(0,0){\circle*{3}}
  \put(10,0){\circle*{3}}
  \put(10,10){\circle*{3}}
  \put(0,10){\circle*{3}}
  \put(10,0){\line(0,1){10}}
\end{picture}
& 1  & $(0,1,0,0)$ &
\begin{picture}(20,20)(0,-2.5)
  \put(0,0){\circle*{3}}
  \put(10,0){\circle*{3}}
  \put(10,10){\circle*{3}}
  \put(0,10){\circle*{3}}
  \put(10,10){\line(-1,0){10}}
  \put(0,10){\line(0,-1){10}}
\end{picture}
& 2  & 
$(0,0,1,1)$ \\ \hline
\begin{picture}(20,20)(0,-2.5)
  \put(0,0){\circle*{3}}
  \put(10,0){\circle*{3}}
  \put(10,10){\circle*{3}}
  \put(0,10){\circle*{3}}
  \put(10,10){\line(-1,0){10}}
\end{picture}
& 1  & $(0,0,1,0)$ &
\begin{picture}(20,20)(0,-2.5)
  \put(0,0){\circle*{3}}
  \put(10,0){\circle*{3}}
  \put(10,10){\circle*{3}}
  \put(0,10){\circle*{3}}
  \put(0,0){\line(1,0){10}}
  \put(10,0){\line(0,1){10}}
  \put(10,10){\line(-1,0){10}}
\end{picture}
& 3  & 
$(1,1,1,0)$ \\ \hline
\begin{picture}(20,20)(0,-2.5)
  \put(0,0){\circle*{3}}
  \put(10,0){\circle*{3}}
  \put(10,10){\circle*{3}}
  \put(0,10){\circle*{3}}
  \put(0,10){\line(0,-1){10}}
\end{picture}
& 1  & $(0,0,0,1)$ &
\begin{picture}(20,20)(0,-2.5)
  \put(0,0){\circle*{3}}
  \put(10,0){\circle*{3}}
  \put(10,10){\circle*{3}}
  \put(0,10){\circle*{3}}
  \put(0,0){\line(1,0){10}}
  \put(10,0){\line(0,1){10}}
  \put(0,10){\line(0,-1){10}}
\end{picture}
& 3  & 
$(1,1,0,1)$ \\ \hline
\begin{picture}(20,20)(0,-2.5)
  \put(0,0){\circle*{3}}
  \put(10,0){\circle*{3}}
  \put(10,10){\circle*{3}}
  \put(0,10){\circle*{3}}
  \put(0,0){\line(1,0){10}}
  \put(10,0){\line(0,1){10}}
\end{picture}
& 2  & $(1,1,0,0)$ &
\begin{picture}(20,20)(0,-2.5)
  \put(0,0){\circle*{3}}
  \put(10,0){\circle*{3}}
  \put(10,10){\circle*{3}}
  \put(0,10){\circle*{3}}
  \put(0,0){\line(1,0){10}}
  \put(10,10){\line(-1,0){10}}
  \put(0,10){\line(0,-1){10}}
\end{picture}
& 3  & 
$(1,0,1,1)$ \\ \hline
\begin{picture}(20,20)(0,-2.5)
  \put(0,0){\circle*{3}}
  \put(10,0){\circle*{3}}
  \put(10,10){\circle*{3}}
  \put(0,10){\circle*{3}}
  \put(0,0){\line(1,0){10}}
  \put(10,10){\line(-1,0){10}}
\end{picture}
& 2  & $(1,0,1,0)$ &
\begin{picture}(20,20)(0,-2.5)
  \put(0,0){\circle*{3}}
  \put(10,0){\circle*{3}}
  \put(10,10){\circle*{3}}
  \put(0,10){\circle*{3}}
  \put(10,0){\line(0,1){10}}
  \put(10,10){\line(-1,0){10}}
  \put(0,10){\line(0,-1){10}}
\end{picture}
& 3  & 
$(0,1,1,1)$ \\ \hline
\begin{picture}(20,20)(0,-2.5)
  \put(0,0){\circle*{3}}
  \put(10,0){\circle*{3}}
  \put(10,10){\circle*{3}}
  \put(0,10){\circle*{3}}
  \put(0,0){\line(1,0){10}}
  \put(0,10){\line(0,-1){10}}
\end{picture}
& 2  & $(1,0,0,1)$ &
\begin{picture}(20,20)(0,-2.5)
  \put(0,0){\circle*{3}}
  \put(10,0){\circle*{3}}
  \put(10,10){\circle*{3}}
  \put(0,10){\circle*{3}}
  \put(0,0){\line(1,0){10}}
  \put(10,0){\line(0,1){10}}
  \put(10,10){\line(-1,0){10}}
  \put(0,10){\line(0,-1){10}}
\end{picture}
& 4  & 
$(1,1,1,1)$ \\ \hline

  \end{tabular}
\caption{Possible configurations at a dual lattice site $x^*$. Active
  bonds are indicated by a solid line, inactive bonds by no line. The
  quantity $w(x^*)=\sum_{i=1}^4 A_i(x^*)$ is the number of active
  bonds surrounding $x^*$.}
\label{tab:1}
\end{table}

Suppose now that we are given an admissible configuration, and we want
to update it to a new admissible configuration. This requires updating
some bonds by changing their state, i.e., $A_i(x^*)\to 1- A_i(x^*)$. A
general set of updated bonds defines a (possibly disconnected) path on
the direct lattice. If this path has an open end, by definition this
means that the end site is touched by a single updated bond. As a
consequence, the parity of the number of active bonds touching this
site would change from odd to even under the update, and the resulting
configuration would not be admissible. The most general admissible
update consists therefore in changing the state of bonds belonging to
a closed (possibly disconnected) path. This is easily seen to be
equivalent to perform consecutive updates on the set of elementary
squares that cover that part of the lattice enclosed by the
path,\footnote{There is an exception: this equivalence does not hold
  when the closed path winds around the lattice, when periodic
  boundary conditions are imposed (see below).}  each elementary
update consisting in changing the state of all the bonds surrounding
each one of the elementary squares. In higher dimensions, this set is
replaced by the elementary squares on a lattice surface having the
path as boundary; any such surface yields the same net update of
bonds.\footnote{When periodic boundary conditions are imposed, one has
  to supplement these updates with those obtained by changing the
  state of all the bonds on straight-line paths winding around the
  lattice. This point is discussed in detail in Section \ref{erg} and
  in \ref{periodic2d} and \ref{ergodic3d}.}

In terms of $A(x^*)$, an elementary update consists in the following
replacement,
\begin{equation}
  \label{eq:s3}
  A(x^*) \to {\cal C} A(x^*) \equiv I - A(x^*)\,,
\end{equation}
where ${\cal C}$ stands for {\it conjugation}, and where we have
introduced the vector $I=(1,1,1,1)$. Under conjugation, $A_i(x^*)\to
1- A_i(x^*)$, $i=1,\ldots 4$; clearly, ${\cal C}^2={\cal I}$, where
${\cal I}$ is the identity, ${\cal I}A(x^*)=A(x^*)$. The variation
$\Delta w(x^*)$ in the number of active bonds under conjugation is
given by
\begin{equation}
  \label{eq:s4}
 \Delta w(x^*)= 
  \sum_{i=1}^4  {\cal C}A_{i}(x^*) - A_i(x^*) = \sum_{i=1}^4  I_i-
  2A_i(x^*)= 2[2- w(x^*)]\,.
\end{equation}

\begin{table}[t]
  \centering
  \begin{tabular}{c|c|c|c|c}
$S(x^*)$ & $A(x^*)$ & $\hat{\cal C} S(x^*)$ & $ {\cal C}A(x^*)$
& $\Delta w(x^*)$                                
 \\ \hline
\begin{picture}(20,20)(0,-2.5)
  \put(0,0){\circle*{3}}
  \put(10,0){\circle*{3}}
  \put(10,10){\circle*{3}}
  \put(0,10){\circle*{3}}
\end{picture}
& $(0,0,0,0)$ &
\begin{picture}(20,20)(0,-2.5)
  \put(0,0){\circle*{3}}
  \put(10,0){\circle*{3}}
  \put(10,10){\circle*{3}}
  \put(0,10){\circle*{3}}
  \put(0,0){\line(1,0){10}}
  \put(10,0){\line(0,1){10}}
  \put(10,10){\line(-1,0){10}}
  \put(0,10){\line(0,-1){10}}
\end{picture}
& $(1,1,1,1)$ 
& 4\\ \hline
\begin{picture}(20,20)(0,-2.5)
  \put(0,0){\circle*{3}}
  \put(10,0){\circle*{3}}
  \put(10,10){\circle*{3}}
  \put(0,10){\circle*{3}}
  \put(0,0){\line(1,0){10}}
\end{picture}
& $(1,0,0,0)$ &
\begin{picture}(20,20)(0,-2.5)
  \put(0,0){\circle*{3}}
  \put(10,0){\circle*{3}}
  \put(10,10){\circle*{3}}
  \put(0,10){\circle*{3}}
  \put(10,0){\line(0,1){10}}
  \put(10,10){\line(-1,0){10}}
  \put(0,10){\line(0,-1){10}}
\end{picture}
& $(0,1,1,1)$ &
2\\ \hline
\begin{picture}(20,20)(0,-2.5)
  \put(0,0){\circle*{3}}
  \put(10,0){\circle*{3}}
  \put(10,10){\circle*{3}}
  \put(0,10){\circle*{3}}
  \put(10,0){\line(0,1){10}}
\end{picture}
& $(0,1,0,0)$ &
\begin{picture}(20,20)(0,-2.5)
  \put(0,0){\circle*{3}}
  \put(10,0){\circle*{3}}
  \put(10,10){\circle*{3}}
  \put(0,10){\circle*{3}}
  \put(0,0){\line(1,0){10}}
  \put(10,10){\line(-1,0){10}}
  \put(0,10){\line(0,-1){10}}
\end{picture}
& $(1,0,1,1)$ 
& 2\\ \hline
\begin{picture}(20,20)(0,-2.5)
  \put(0,0){\circle*{3}}
  \put(10,0){\circle*{3}}
  \put(10,10){\circle*{3}}
  \put(0,10){\circle*{3}}
  \put(10,10){\line(-1,0){10}}
\end{picture}
& $(0,0,1,0)$ &
\begin{picture}(20,20)(0,-2.5)
  \put(0,0){\circle*{3}}
  \put(10,0){\circle*{3}}
  \put(10,10){\circle*{3}}
  \put(0,10){\circle*{3}}
  \put(0,0){\line(1,0){10}}
  \put(10,0){\line(0,1){10}}
  \put(0,10){\line(0,-1){10}}
\end{picture}
& $(1,1,0,1)$ &
2\\ \hline
\begin{picture}(20,20)(0,-2.5)
  \put(0,0){\circle*{3}}
  \put(10,0){\circle*{3}}
  \put(10,10){\circle*{3}}
  \put(0,10){\circle*{3}}
  \put(0,10){\line(0,-1){10}}
\end{picture}
& $(0,0,0,1)$ &
\begin{picture}(20,20)(0,-2.5)
  \put(0,0){\circle*{3}}
  \put(10,0){\circle*{3}}
  \put(10,10){\circle*{3}}
  \put(0,10){\circle*{3}}
  \put(0,0){\line(1,0){10}}
  \put(10,0){\line(0,1){10}}
  \put(10,10){\line(-1,0){10}}
\end{picture}
& $(1,1,1,0)$ &
2 \\ \hline
\begin{picture}(20,20)(0,-2.5)
  \put(0,0){\circle*{3}}
  \put(10,0){\circle*{3}}
  \put(10,10){\circle*{3}}
  \put(0,10){\circle*{3}}
  \put(0,0){\line(1,0){10}}
  \put(10,0){\line(0,1){10}}
\end{picture}
& $(1,1,0,0)$ &
\begin{picture}(20,20)(0,-2.5)
  \put(0,0){\circle*{3}}
  \put(10,0){\circle*{3}}
  \put(10,10){\circle*{3}}
  \put(0,10){\circle*{3}}
  \put(10,10){\line(-1,0){10}}
  \put(0,10){\line(0,-1){10}}
\end{picture}
& $(0,0,1,1)$ &
0\\ \hline
\begin{picture}(20,20)(0,-2.5)
  \put(0,0){\circle*{3}}
  \put(10,0){\circle*{3}}
  \put(10,10){\circle*{3}}
  \put(0,10){\circle*{3}}
  \put(0,0){\line(1,0){10}}
  \put(10,10){\line(-1,0){10}}
\end{picture}
& $(1,0,1,0)$ & 
\begin{picture}(20,20)(0,-2.5)
  \put(0,0){\circle*{3}}
  \put(10,0){\circle*{3}}
  \put(10,10){\circle*{3}}
  \put(0,10){\circle*{3}}
  \put(10,0){\line(0,1){10}}
  \put(0,10){\line(0,-1){10}}
\end{picture}
& $(0,1,0,1)$ &
0\\ \hline
\begin{picture}(20,20)(0,-2.5)
  \put(0,0){\circle*{3}}
  \put(10,0){\circle*{3}}
  \put(10,10){\circle*{3}}
  \put(0,10){\circle*{3}}
  \put(0,0){\line(1,0){10}}
  \put(0,10){\line(0,-1){10}}
\end{picture}
& $(1,0,0,1)$ &
\begin{picture}(20,20)(0,-2.5)
  \put(0,0){\circle*{3}}
  \put(10,0){\circle*{3}}
  \put(10,10){\circle*{3}}
  \put(0,10){\circle*{3}}
  \put(10,0){\line(0,1){10}}
  \put(10,10){\line(-1,0){10}}
\end{picture}
& $(0,1,1,0)$ &
0 \\ \hline
  \end{tabular}
\caption{Transformation of configurations under conjugation ${\cal
    C}$. Acted upon a graph, we denote the transformation with a hat.
  The quantity $\Delta w(x^*)$ is the variation in the number of
  active bonds when passing from $S(x^*)$ to $\hat{\cal C}S(x^*)$. As
  $\hat{\cal C}$ is an involution, $\hat{\cal C}[\hat{\cal
      C} S(x^*)] =S(x^*)$, so the conjugate of a configuration in
  column 3 is the corresponding configuration in column 1, and the
  variation of the number of active bonds changes sign when passing
  from a configuration in column 3 to the corresponding configuration
  in column 1.  }
\label{tab:2}
\end{table}
In Table 2 we show the pairs of configurations of an elementary square
connected by conjugation, together with $\Delta w(x^*)$.  These
updating steps can be applied independently to all the sites of the
dual lattice, and are clearly reversible. We have now the ingredients
to set up the first version of a Metropolis algorithm (Algorithm
\ref{met1}).

\smallskip

\begin{algorithm}[h!]
\caption{}
\label{met1}
\begin{enumerate}
\item At a given site $x^*$ of the dual lattice, compute $\Delta
  w(x^*)$ corresponding to a conjugation step.
\item If $\Delta w(x^*)\le 0$, accept the step.
\item If $\Delta w(x^*)> 0$, take a random number $r\in[0,1]$. If
  $r\le \tanh(|F|)^{\Delta w(x^*)}$, accept the step, otherwise
  reject it (note that in this case $\tanh(|F|)^{\Delta w(x^*)} \le 1$). 
\item Repeat the procedure for all the dual lattice sites.
\end{enumerate}
\end{algorithm}
\noindent
It is easy to see that this algorithm satisfies detailed balance. The
only question remaining is that of the ergodicity of the algorithm.

\subsection{Ergodicity}
\label{erg}

Concerning ergodicity the situation is slightly different for open or
periodic boundary conditions.

The simplest case is that of open boundaries. In this case we can
prove (\ref{open2d}) that all the admissible configurations
can be transformed to the same configuration through a sequence of
conjugation moves. As these transformations are reversible, any
admissible configuration is connected to all the others. Therefore in
this case algorithm \ref{met1} is ergodic and can be used as is to
simulate the system.

The case of periodic boundary conditions is slightly more
complicated. One can see that in this case, the total number of
vertical (respectively horizontal) active bonds modulo 2 on any given
row (respectively, column) of the dual lattice is conserved under the
updating moves, and moreover is the same for any row (column). Calling
these numbers vertical and horizontal parity, $P_{V}$ and $P_{H}$,
respectively, this defines four different ``sectors'' of admissible
configurations, classified by parities ($P_{V}, P_{H}$) $\in$
$\{(0,0), (1, 0), (0, 1), (1,1)\}$. It can be shown that the updating
moves are ergodic within each sector separately (\ref{periodic2d}).  Therefore we have to modify slightly algorithm
\ref{met1} to obtain an ergodic algorithm (algorithm \ref{met2}).

\smallskip

\begin{algorithm}[h!]
\caption{}
\label{met2}
\begin{enumerate}
\item Start from a configuration of parity, say,  $(0,0)$.
\item After a certain number of sweeps through the whole lattice,
  using algorithm \ref{met1}, propose a change of the parity $P_H$ of
  the configuration, by proposing the inversion of the state of all
  the horizontal bonds on a row $i=i_0$, $j=1,\ldots L$, of the direct
  lattice, chosen randomly, so (possibly) passing to a configuration
  of parity $(0,1)$ with probability $\tanh(|F|)^{\Delta \tilde w}$,
  where
  \begin{equation}
    \label{eq:algeq1h}
 \Delta\tilde w=
  \sum_{j=1}^L \left({\cal C}B_H(i_0,j) - B_H(i_0,j)\right) = 
L-2\sum_{j=1}^L B_H(i_0,j)   \,,
  \end{equation}
and we have denoted $B_H(i,j)=A_1(i,j)$, and ${\cal
  C}B_H(i,j)=1-B_H(i,j)$.
\item After the same number of sweeps through the whole lattice, (try
  to) change the parity $P_V$ of the configuration by proposing the
  inversion of the state of all the vertical bonds on a column
  $j=j_0$, $i=1,\ldots L$, of the direct lattice, chosen randomly, so
  (possibly) passing to a configuration of parity $(1,1)$, if the
  proposal in the previous point has been accepted, or $(1,0)$, if the
  proposal in the previous point has been rejected, with probability
  $\tanh(|F|)^{\Delta \tilde w}$, where now
  \begin{equation}
    \label{eq:algeq1v}
 \Delta\tilde w=
  \sum_{i=1}^L \left( {\cal C}B_V(i,j_0) - B_V(i,j_0) \right) = 
L-2\sum_{i=1}^L
  B_V(i,j_0)   \,, 
  \end{equation}
and we have denoted $B_V(i,j)=A_4(i,j)$, and ${\cal
  C}B_V(i,j)=1-B_V(i,j)$.
\item Iterate the procedure. 
\end{enumerate}
\end{algorithm}
\noindent

The case of higher dimensionality is a straightforward generalization
of the algorithms presented here, obtained by applying algorithm
\ref{met1} to all the elementary squares of the lattice, and by
taking into account that for periodic boundary conditions there are
now $2^D$ ``sectors'', classified by the $D$ parities defined in analogy
to the $2D$ case.

\section{Numerical results}
\label{numerical}

We have performed numerical simulations of both the $2D$ and the $3D$
models using the algorithms discussed previously.

First of all, we tested the algorithms in the case of periodic
boundary conditions. The acceptance rate of the global update varies
widely with both the coupling and the size of the system. For example,
for the $2D$ model at $F = -1.0$ the acceptance rate drops from about
$60\%$ at $L = 16$ to about $30\%$ at $L = 64$; for a smaller
coupling, $F = -0.6$, the acceptance rate drops from $27\%$ at $L =
16$ to $3\%$ at $L = 64$.\footnote{For large volumes and small
  couplings the global acceptance rate is so small that our
  simulations are effectively confined to one sector. However the
  difference between sectors is a boundary effect, and we expect, on
  general thermodynamic grounds, that it should vanish in the large
  volume limit. This is strongly supported by the very precise
  agreement, for all values of the coupling, between our simulations
  and the exact results in the $2D$ case.}

\begin{figure}[h!]
\begin{center}
\includegraphics[scale=0.35]{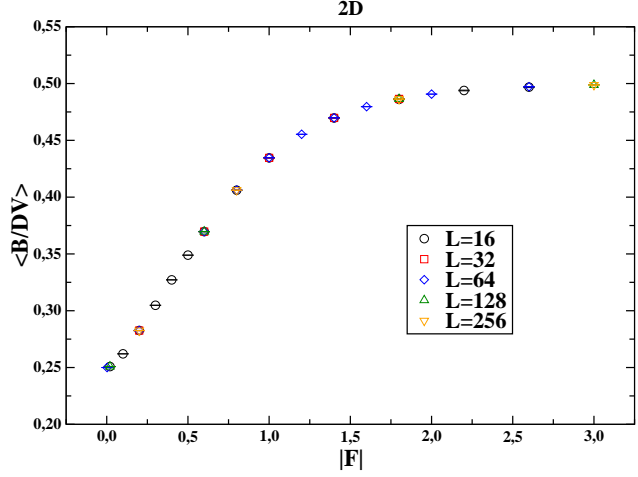}

\includegraphics[scale=0.35]{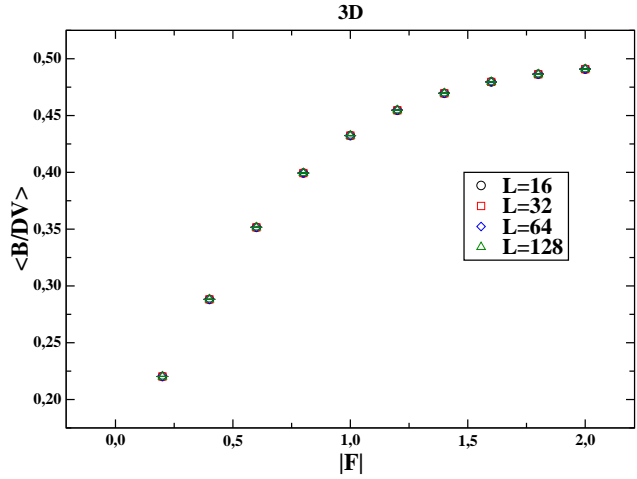}
\end{center}
\caption{Average number of active bonds normalized by the total number
  of bonds as a function of $|F|$ for different lattice sizes $L$ for
  the $2D$ model (top) and $3D$ model (bottom).}
\label{numero_medio_dimeri}
\end{figure}

Then we checked that the bounds on the number of active bonds in each
configuration of our system ( \ref{extremebonds}) and also on the
average of this quantity ( \ref{averagebonds}) were respected. For
simplicity we only show in Figs.~(\ref{numero_medio_dimeri}) the
average number of active bonds.

Then we calculated the correlation functions~\eqref{eq:14bis} as well
as the energy density~\eqref{equation:energy} and the specific
heat~\eqref{equation:specific} both for the two-dimensional and
three-dimensional systems. We have evaluated these quantities for
different values of the coupling $F$ and various volumes, with lattice
sizes ranging from $L=16$ to $L=1024$.  Simulations were done
collecting $100$k measurements for each value of $F$.  We discarded
between $10$k and $20$k configurations at the beginning of each run in
order to ensure thermalization.  The data analysis was done using the
jackknife method over bins at different blocking levels.

\begin{figure}[p]
\begin{center}
\includegraphics[scale=0.35]{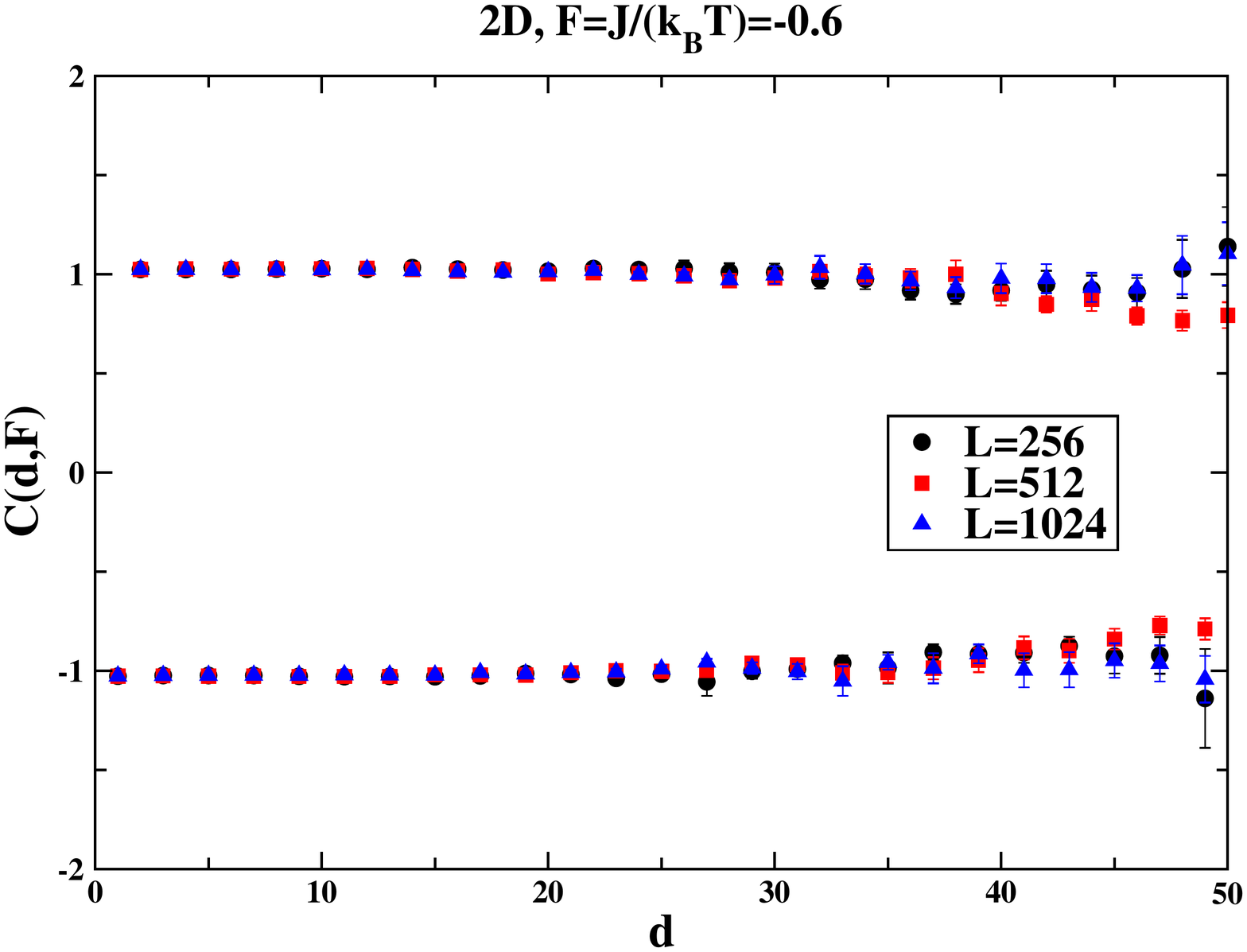}
\includegraphics[scale=0.35]{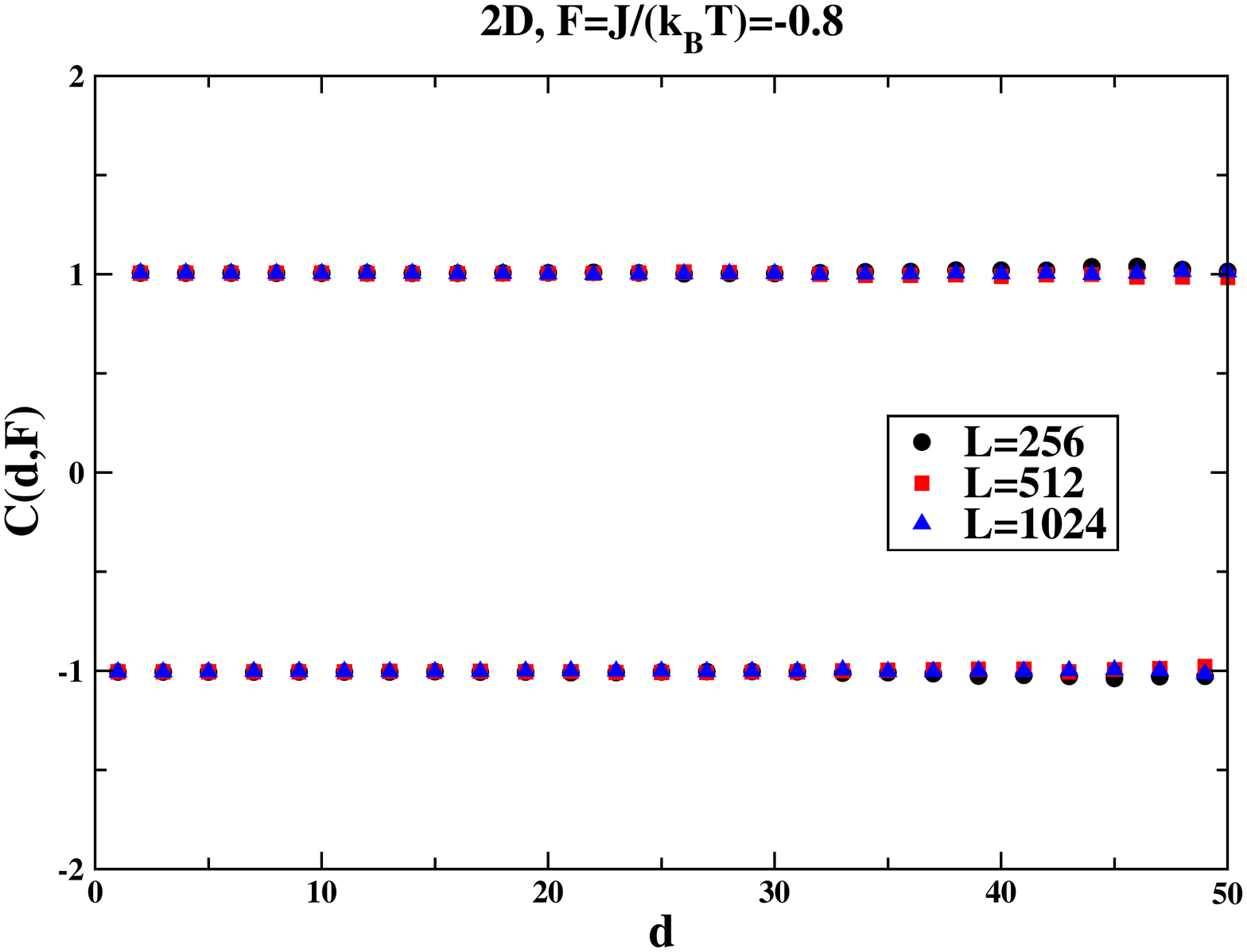}
\end{center}
\caption{Dependence of the correlation functions on the distance $d$
  for different values of the coupling $F$ and for various lattice
  sizes $L$ in the $2D$ model.}
\label{2D_corr1}
\end{figure}

\begin{figure}[p]
\begin{center}
\includegraphics[scale=0.35]{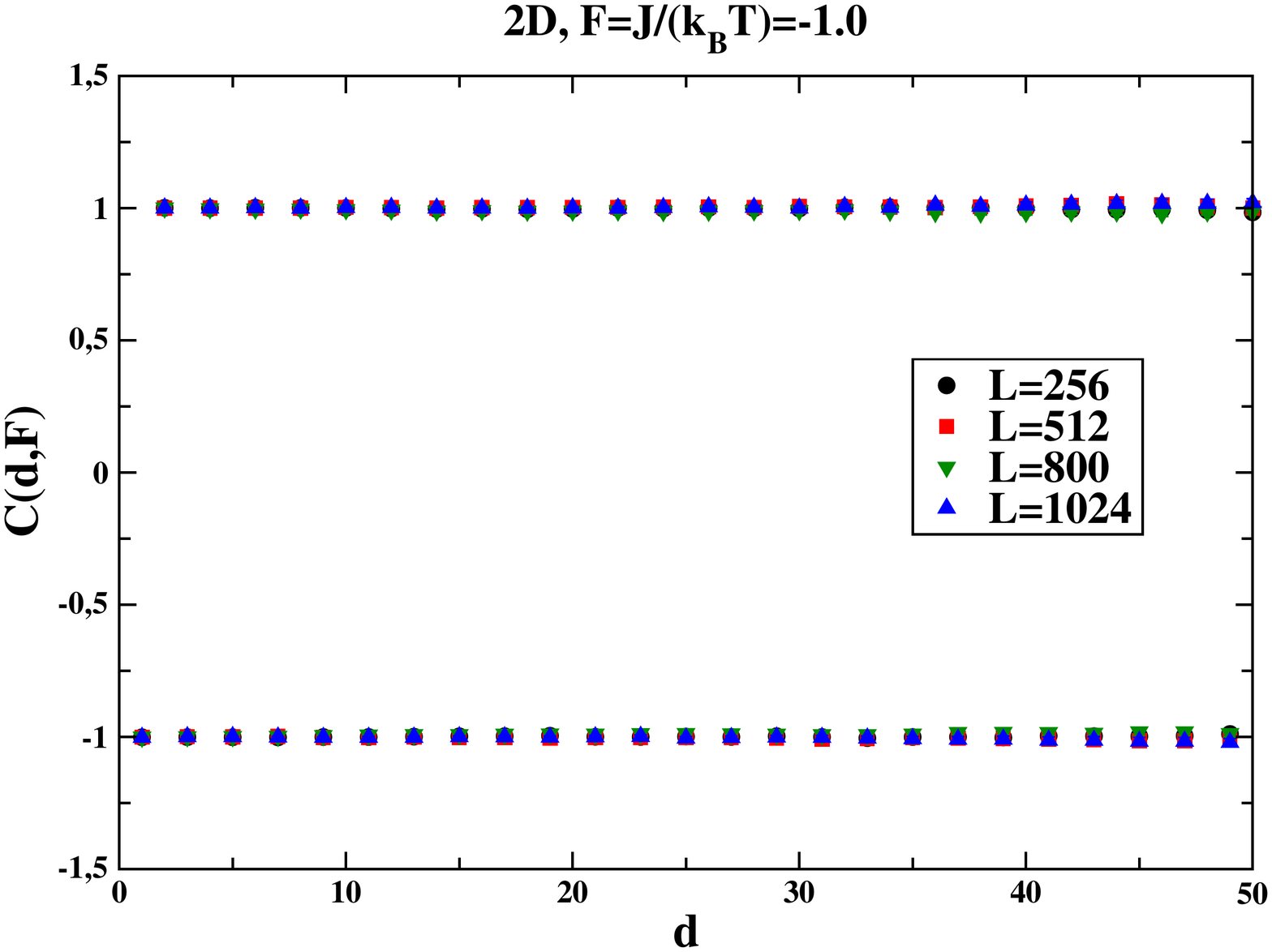}
\includegraphics[scale=0.35]{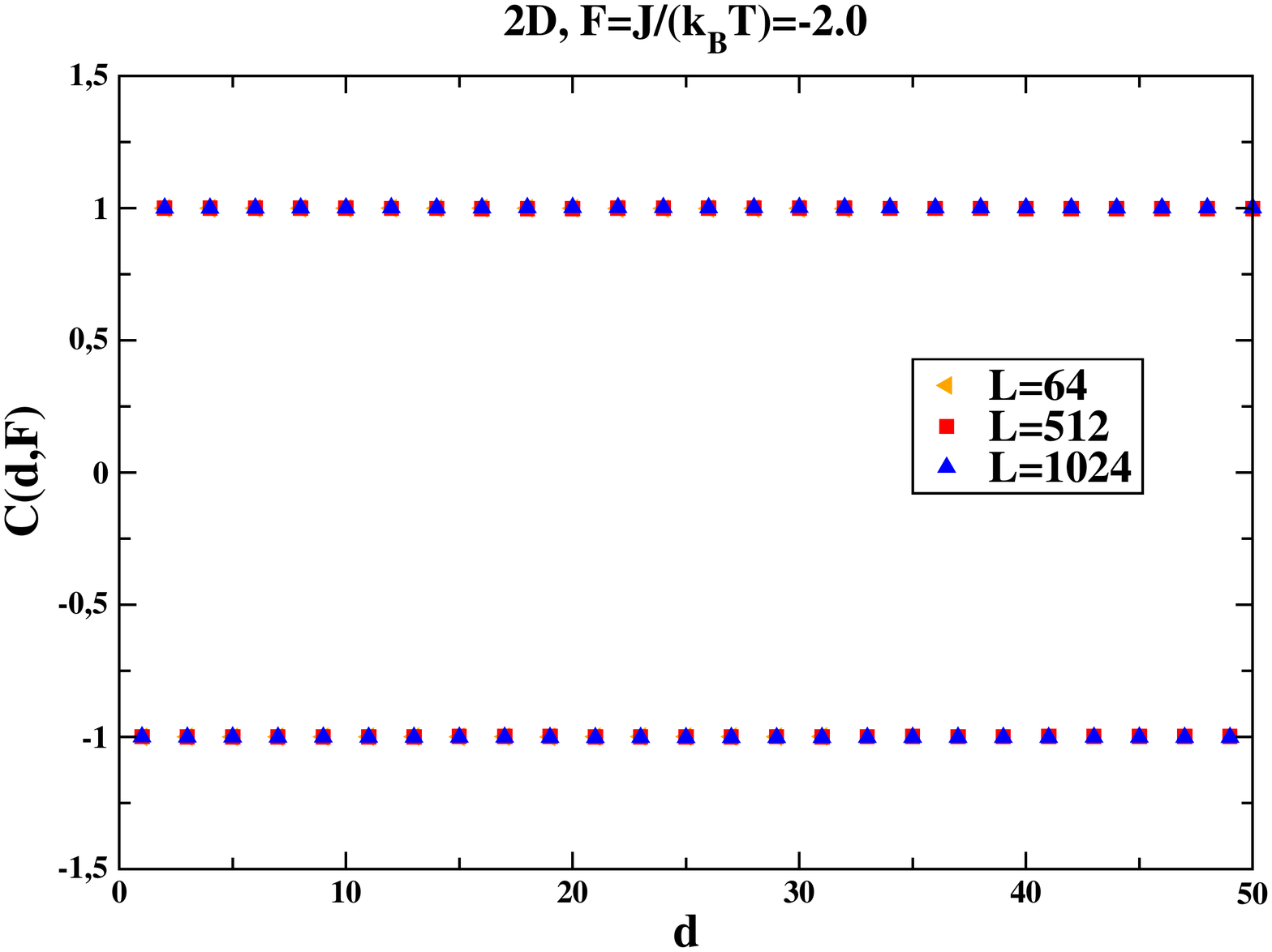}
\end{center}
\caption{Dependence of the correlation functions on the distance $d$
  for different values of the coupling $F$ and for various lattice
  sizes $L$ in the $2D$ model (continued).}
\label{2D_corr2}
\end{figure}

\begin{figure}[p]
\begin{center}
\vspace{-1cm}
\includegraphics[scale=0.30]{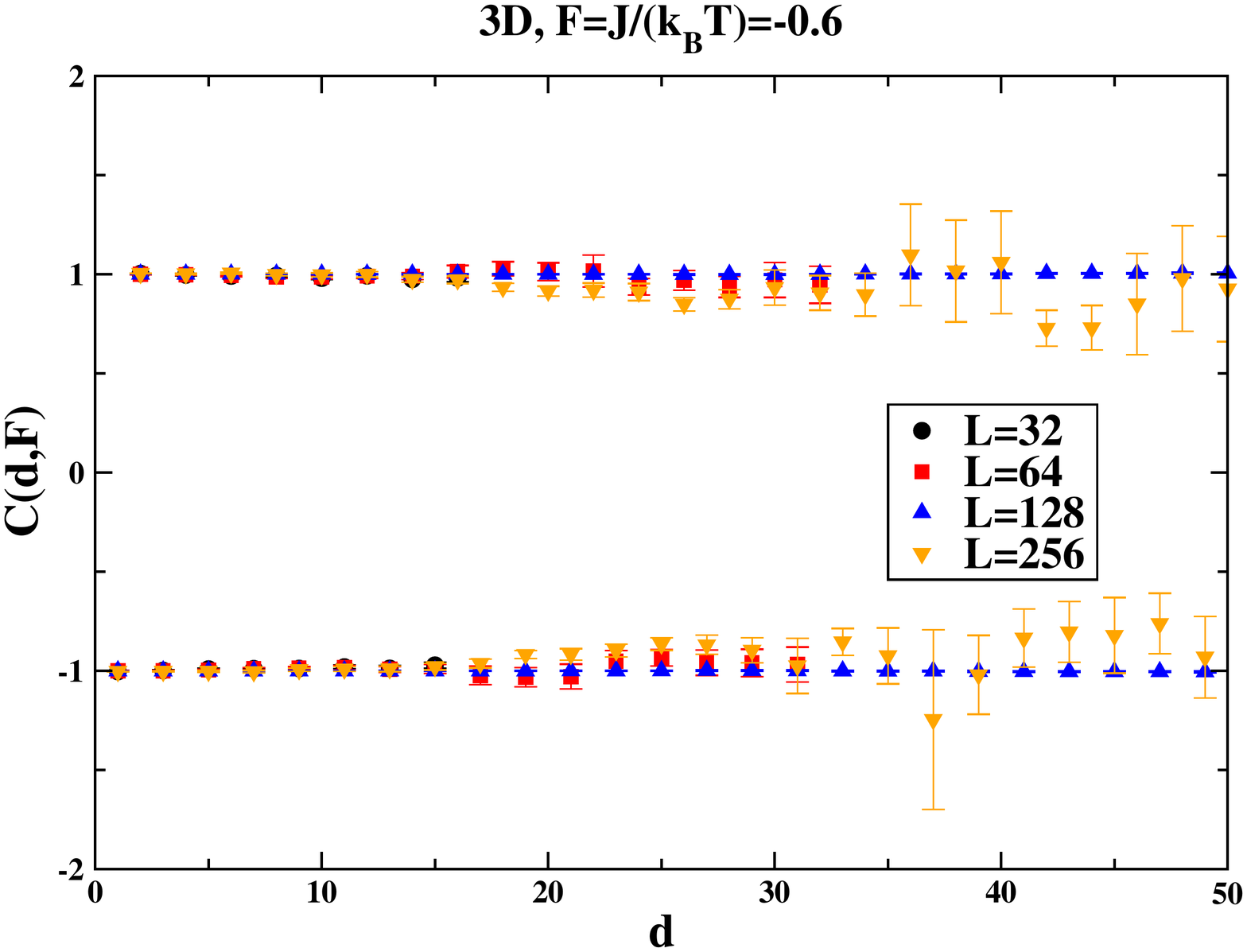}
\includegraphics[scale=0.30]{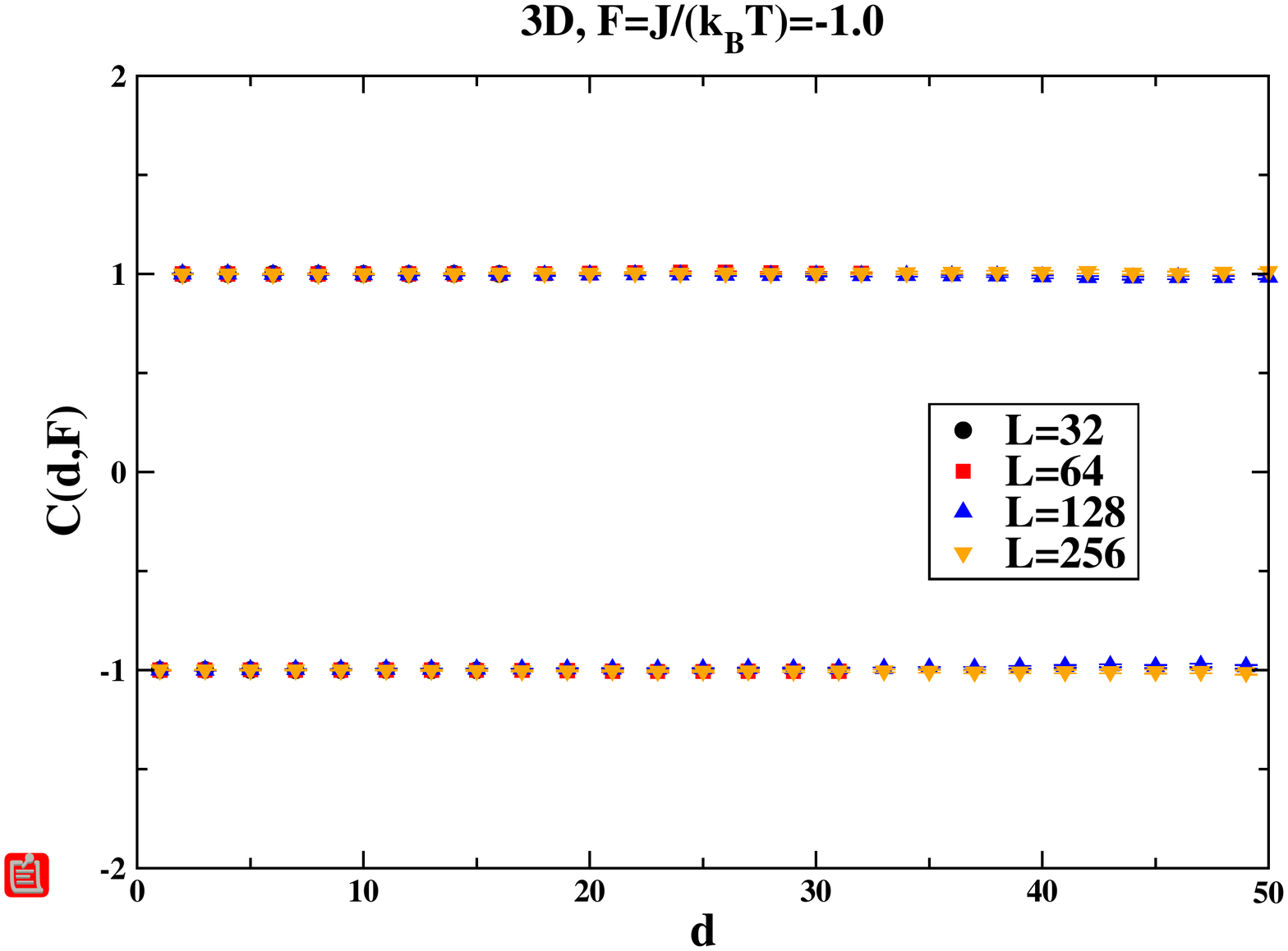}
\includegraphics[scale=0.30]{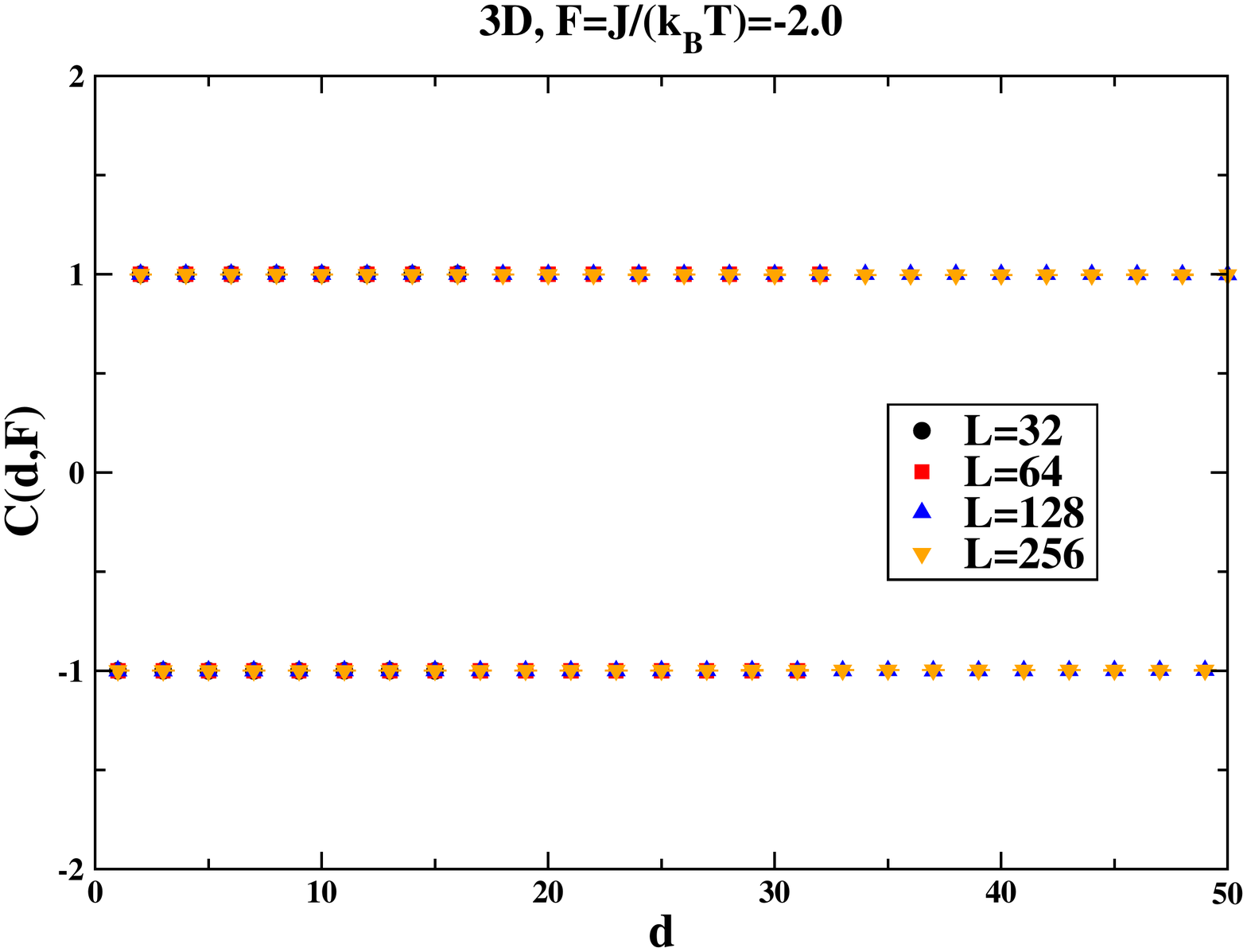}
\end{center}
\caption{Dependence of the correlation functions on the distance $d$
  for different values of the coupling $F$ and for various lattice
  sizes $L$ in the $3D$ model.}
\label{3D_corr}
\end{figure}

\begin{figure}[p]
\begin{center}
\vspace{-1.5cm}
\includegraphics[scale=0.45]{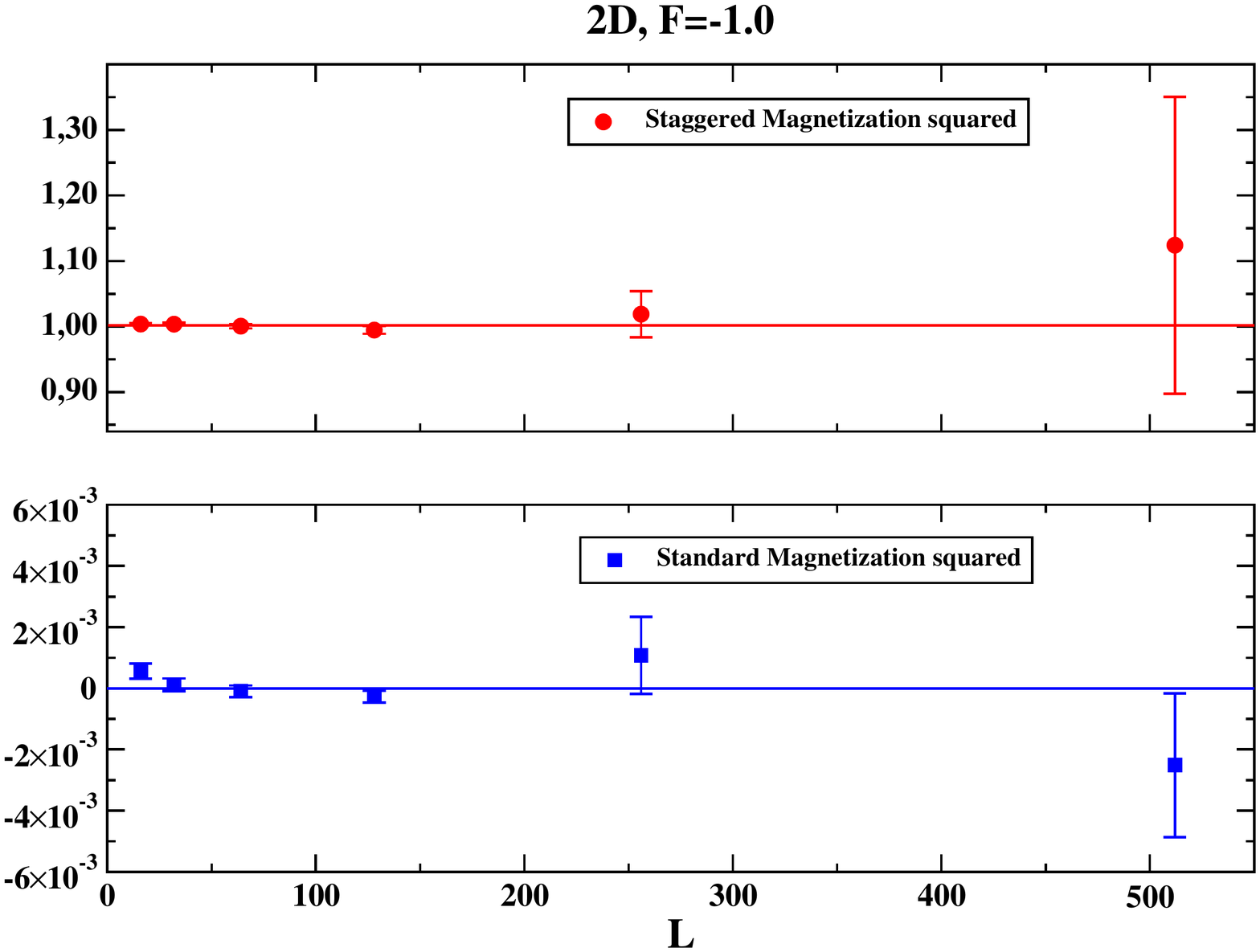}
\includegraphics[scale=0.45]{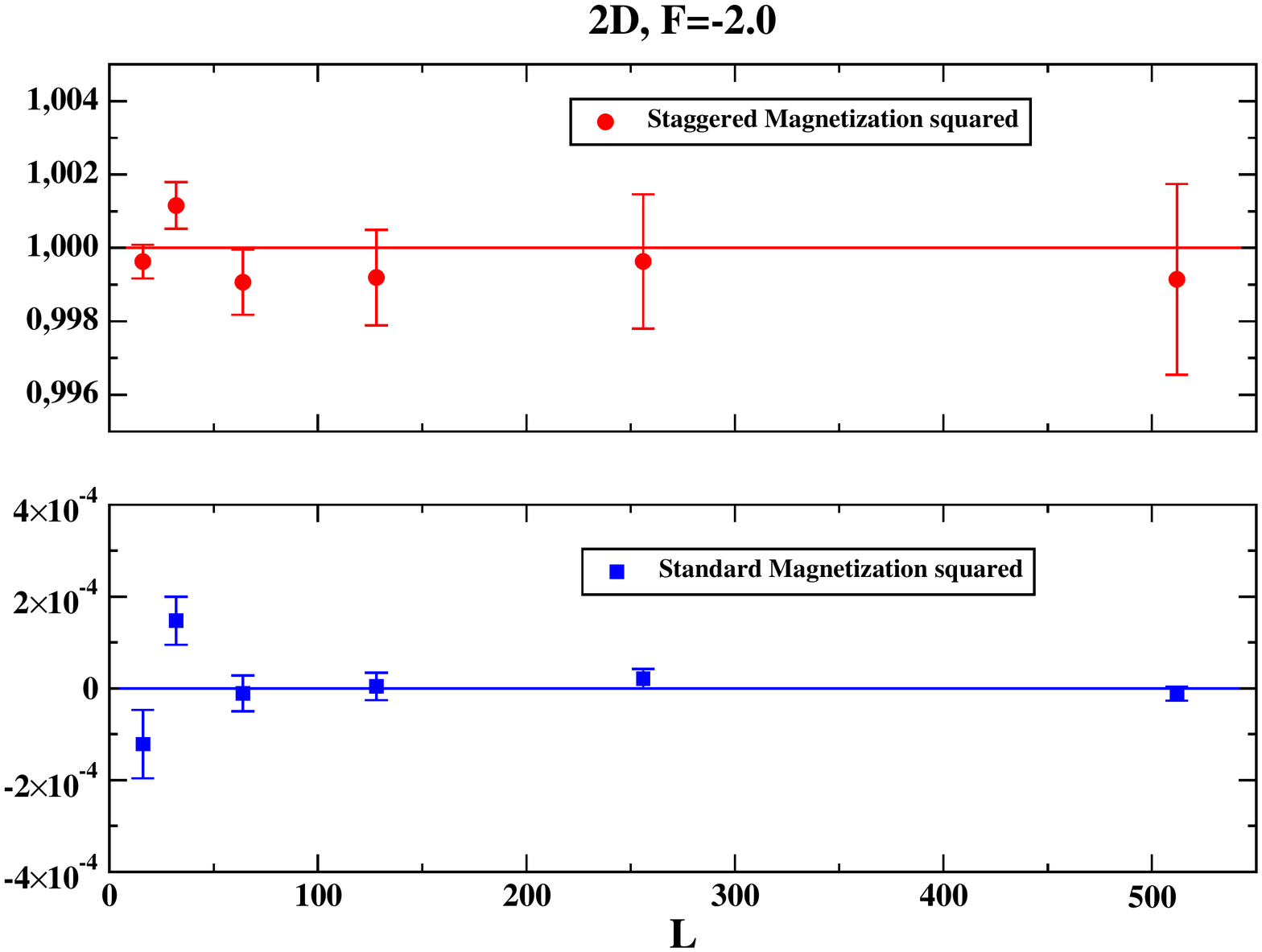}

\end{center}
\caption{Standard and staggered magnetization squared for two values
  of the coupling $F$ for the $2D$ model. The red line indicates the
  analytical prediction (Eq.~\eqref{staggered_magn}, \ref{analytic}),
  whereas the blue line is the zero value.}
\label{ms_2D}
\end{figure}

\begin{figure}[p]
\begin{center}
\vspace{-1.5cm}
\includegraphics[scale=0.45]{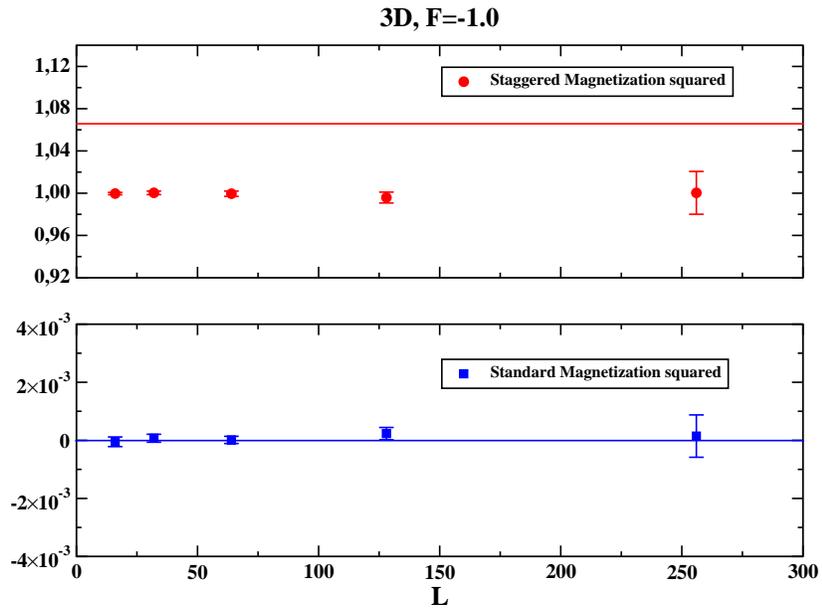}
\includegraphics[scale=0.45]{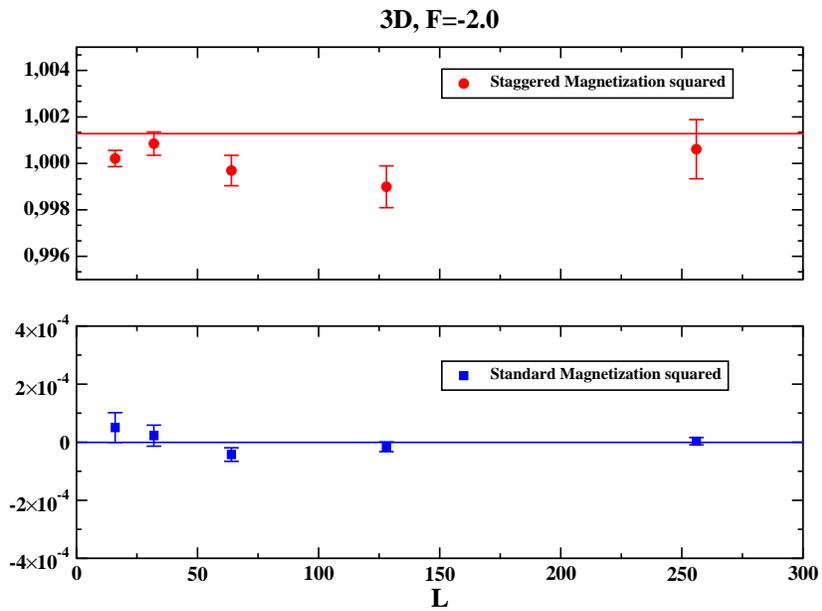}

\end{center}
\caption{Standard and staggered magnetization squared for two values
  of the coupling $F$ for the $3D$ model. The red line indicates the
  mean-field prediction \cite{rif1}, whereas the blue line is the zero
  value.}
\label{ms_3D}
\end{figure}

\begin{figure}[p]
\begin{center}
\includegraphics[scale=0.35]{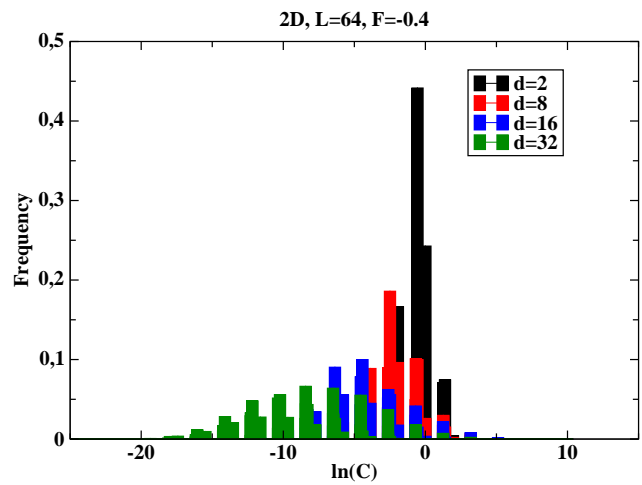}
\includegraphics[scale=0.35]{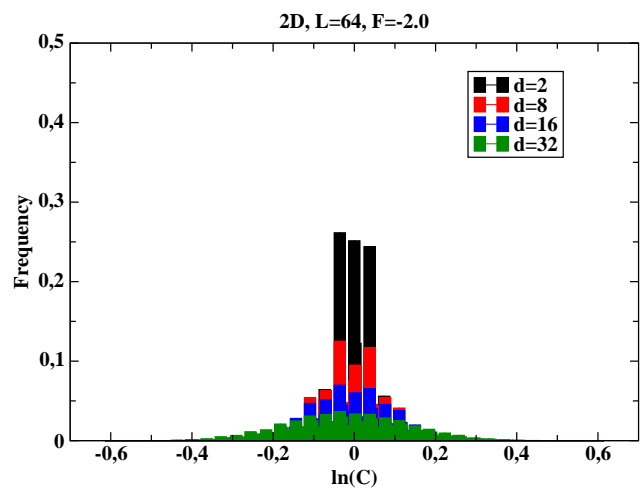}
\end{center}
\caption{Behavior of the probability distributions of the logarithm of
  the correlators in the $2D$ model for two different values of the
  coupling $F$.}
\label{distribuzioni}
\end{figure}

\begin{figure}[p]
\begin{center}
\includegraphics[scale=0.75]{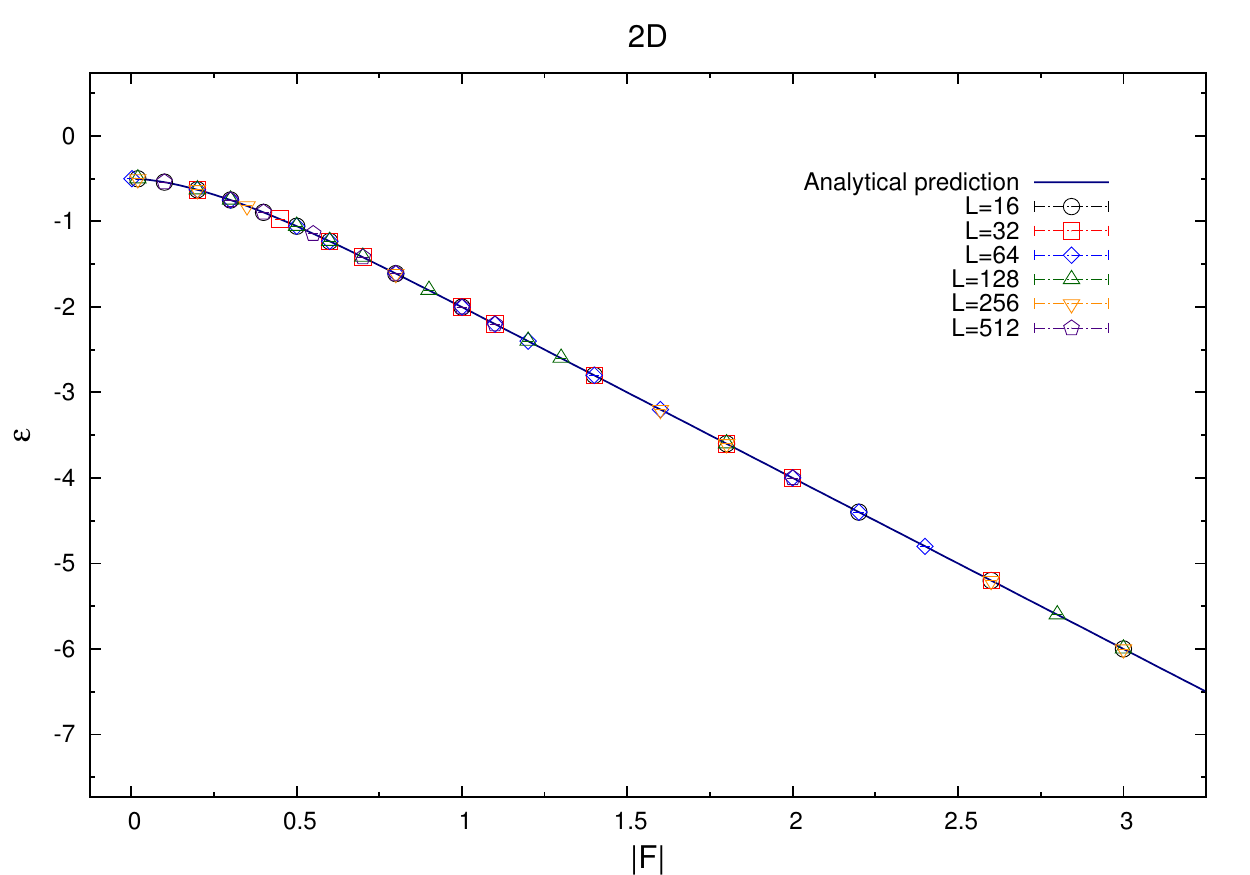}

\vspace{1cm}

\includegraphics[scale=0.75]{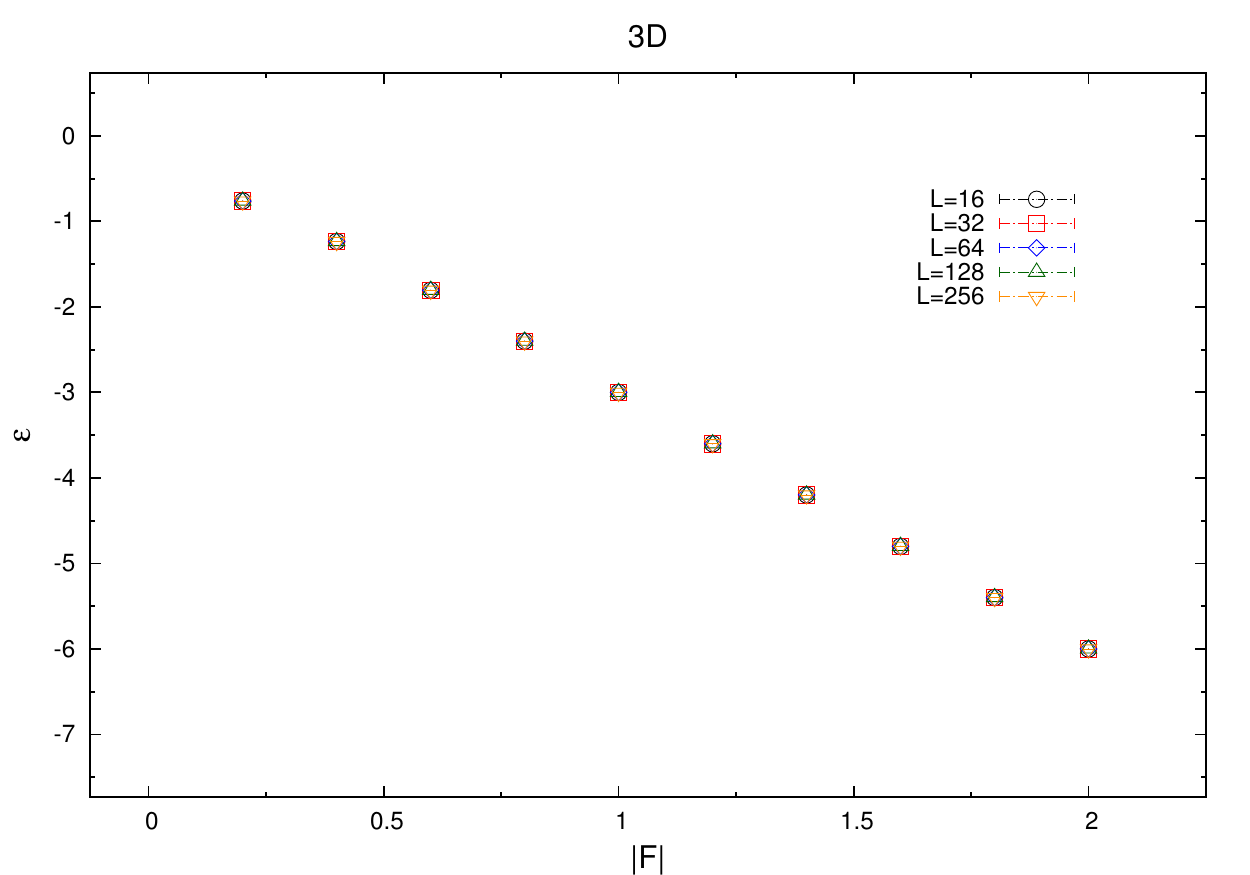}
\end{center}
\caption{Energy density $\varepsilon$ as a function of $|F|$ for
  various lattice sizes $L$ for the $2D$ model (top) and $3D$ model
  (bottom). The analytical result for the $2D$ model is also plotted
  for comparison.}
\label{Energy}
\end{figure}

\begin{figure}[p]
\begin{center}
\includegraphics[scale=0.75]{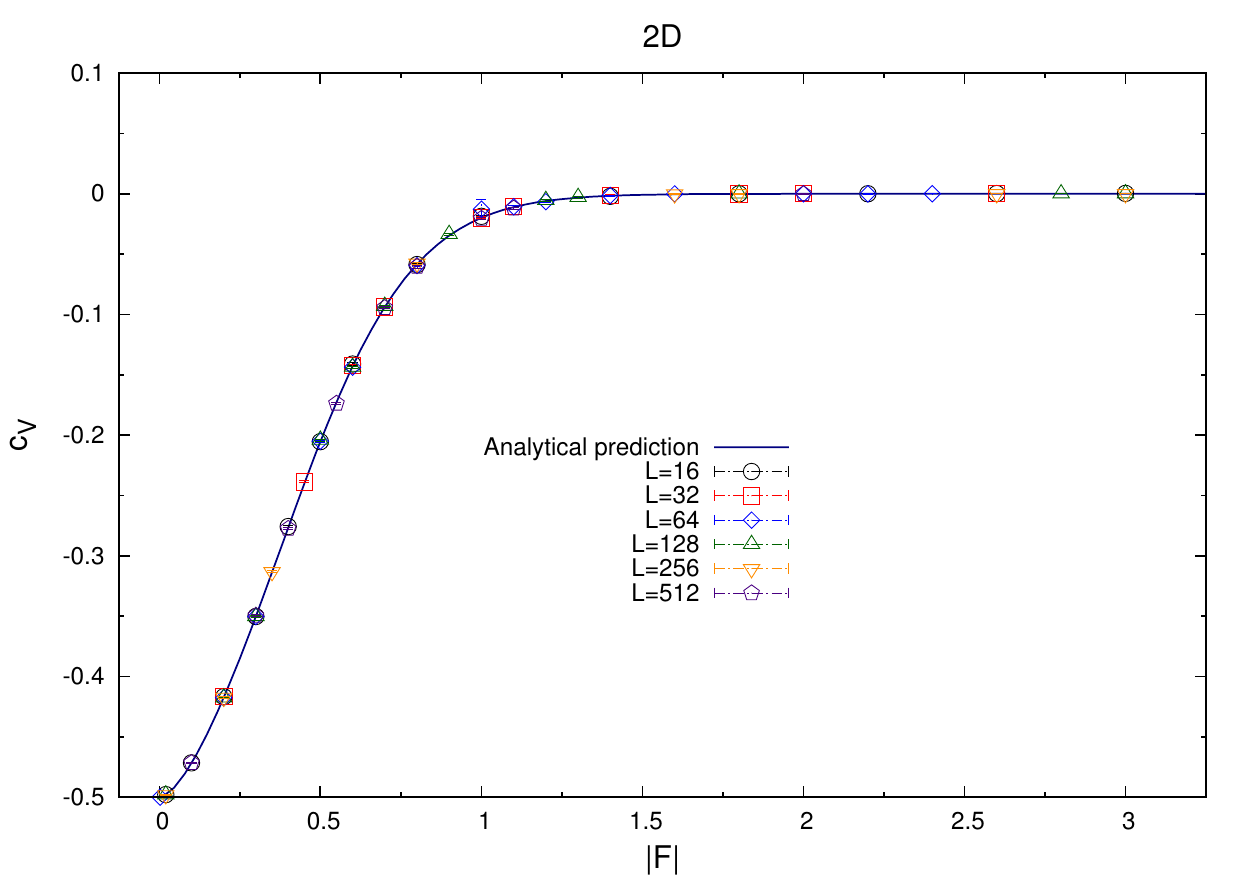}


\includegraphics[scale=0.75]{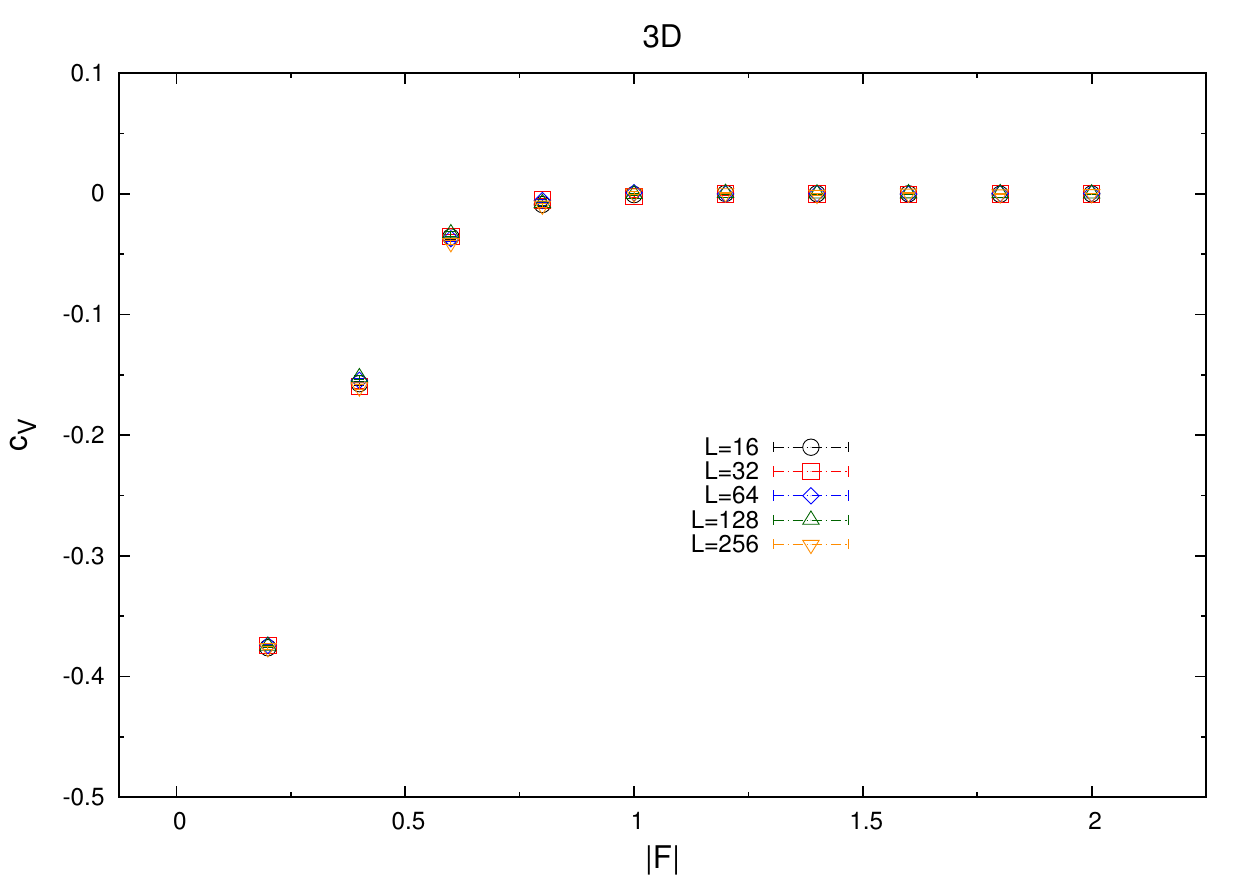}
\end{center}
\caption{Specific heat $c_{V}$ as a function of $|F|$ for various lattice
  sizes $L$ for the $2D$ model (top) and the $3D$ model (bottom). The
  analytical result for the $2D$ model is also plotted for
  comparison.}
\label{Specific}
\end{figure}

In Figs.~(\ref{2D_corr1}),~(\ref{2D_corr2}) and~(\ref{3D_corr}) we
show how the correlation functions depend on the distance $d$, both
for the $2D$ and $3D$ models, choosing different antiferromagnetic
couplings $F < 0$ and varying the lattice volume $V$. In
Fig.~(\ref{ms_2D}) we show the staggered and the standard
magnetization squared in the $2D$ model for two values of the coupling
$F$, together with a solid line indicating the analytical result. As
can be seen the staggered magnetization squared is always different
from zero while the standard magnetization vanishes for all values of
the coupling $F$, and the results are in perfect agreement with the
analytical solution of Refs.~\cite{Lee,Matveev,Wu}. In
Fig.~(\ref{ms_3D}) we show the corresponding results for the $3D$
model at the same values of the coupling, as well as the mean-field
prediction for the staggered magnetization obtained in
Ref.~\cite{rif1}. Again the standard magnetization vanishes, whereas
the staggered one does not, and its value is close to the mean-field
prediction for large values of the coupling $|F|$. Therefore, despite
the vanishing of the standard magnetization, the $Z_2$ symmetry of the
system is spontaneously broken both in the $2D$ and in the $3D$
models. We can also notice an apparent decrease of $C(d,F)$ at large
$d$ for small values of the coupling $F$ both for the $2D$ and $3D$
models.  This is due to the heavy-tailed distributions of the
correlators, which are also responsible for the noisy behavior seen at
large $d$. In Fig.~(\ref{distribuzioni}) we show the probability
distributions of the logarithm of the correlators for a lattice size
$L=64$ and for $F=-0.4$ and $F=-2.0$. Clearly we notice that for a
small coupling $|F|$ the values are spread in a wider range than for
$F=-2.0$; also a long tail develops for large distances $d$, thus
making more difficult a precise evaluation of the
correlators.\footnote{This is the reason why we have made no attempt
  to calculate the staggered magnetization for couplings smaller than
  $|F| = 1.0$.} Finally, in Figs.~(\ref{Energy}) and~(\ref{Specific})
we show the energy density $\varepsilon$ and the specific heat $c_{V}$
as a function of $|F|$ for various lattice sizes $L$. The solid lines
present in the figures for the $2D$ model are the analytical results
for these quantities~\cite{Matveev}. We can see that the energy
density and the specific heat do not show any sign of singular
behavior in $F$ for the range of couplings studied.

\section{Conclusions}
\label{conclusions}

In this paper we studied the $2D$ and $3D$ antiferromagnetic Ising
model with a ``topological'' $\theta$-term at $\theta = \pi$. For this
model we introduced a new geometric algorithm free from the sign
problem.

The numerical part of the work has been devoted to testing the
algorithm for the two-dimensional model against known analytical
results, with which we obtain perfect agreement, and then afterwards
to study the three-dimensional system. Our findings strongly support
the scenario that, despite the vanishing of the standard
magnetization, the staggered magnetization is non-zero for all $D \ge
2$, and therefore the $Z_{2}$ symmetry is spontaneously broken for all
values of $F$ at $\theta=\pi$.

It would be interesting to study whether it is possible to introduce a
``worm'' in our algorithm, in the spirit of \cite{worm1}. This could
change the dynamics of the system; in particular an implementation
that allows the worm to wind through the lattice might be able to
tunnel between the different parity sectors. We leave the study of
such a possibility for a future work.

\section{Acknowledgments}
The work was funded by MICINN (under grant FPA2012-35453 and
FPA2009-09638), DGIID-DGA (grant 2007-E24/2), and by the EU under
ITN-STRONGnet (PITN-GA-2009-238353). EF was supported by the MICINN
Ramon y Cajal program. MG is supported by the Hungarian Academy of
Sciences under ``Lend\"ulet'' grant No. LP2011-011, and partially by
MICINN under the CPAN project CSD2007-00042 from the
Consolider-Ingenio2010 program.

\appendix

\section{Ergodicity}
\label{ergodicity}

\subsection{Open boundary 
conditions for the $2D$ model}
\label{open2d}

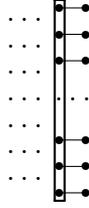
\begin{figure}[t]
  \centering
  \vspace{0.5cm}
  \begin{picture}(80,80)(-40,-15)
  \put(0,0){\circle*{3}}
  \put(10,0){\circle*{3}}
  \put(10,10){\circle*{3}}
  \put(0,10){\circle*{3}}

  \put(10,20){\circle*{3}}
  \put(0,20){\circle*{3}}

   \put(-1.75,35){\ldots}
 \put(10,50){\circle*{3}}
  \put(0,50){\circle*{3}}
 \put(10,60){\circle*{3}}
  \put(0,60){\circle*{3}}
\put(10,70){\circle*{3}}
  \put(0,70){\circle*{3}}

  \put(0,0){\line(1,0){10}}
  \put(0,10){\line(1,0){10}}
  \put(0,20){\line(1,0){10}}
  \put(0,50){\line(1,0){10}}
  \put(0,60){\line(1,0){10}}
  \put(0,70){\line(1,0){10}}

  \put(-20,5){\ldots}
  \put(-20,15){\ldots}
  \put(-20,25){\ldots}
  \put(-20,35){\ldots}
  \put(-20,45){\ldots}
  \put(-20,55){\ldots}
  \put(-20,65){\ldots}

\thicklines

  \put(-1.75,-3){\line(0,1){76}}
  \put(2,-3){\line(0,1){76}}
  \put(-2.15,-3){\line(1,0){4.5}}
  \put(-2.15,73){\line(1,0){4.5}}

\end{picture}
  \caption{Right-most column of the dual lattice (using open boundary
    conditions) after the first step of reduction. The state of the
    vertical bonds inside the rectangle is not specified.}
  \label{fig:2}
\end{figure}

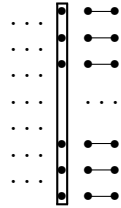
\begin{figure}[t]
  \centering
  \begin{picture}(80,80)(-40,-15)
  \put(0,0){\circle*{3}}
  \put(10,0){\circle*{3}}
  \put(10,10){\circle*{3}}
  \put(0,10){\circle*{3}}

  \put(10,20){\circle*{3}}
  \put(0,20){\circle*{3}}

  \put(-1.75,35){\ldots}
 \put(10,50){\circle*{3}}
  \put(0,50){\circle*{3}}
 \put(10,60){\circle*{3}}
  \put(0,60){\circle*{3}}
\put(10,70){\circle*{3}}
  \put(0,70){\circle*{3}}

  \put(-10,0){\circle*{3}}
  \put(-10,10){\circle*{3}}
  \put(-10,20){\circle*{3}}

 \put(-10,50){\circle*{3}}
 \put(-10,60){\circle*{3}}
 \put(-10,70){\circle*{3}}

  \put(0,0){\line(1,0){10}}
  \put(0,10){\line(1,0){10}}
  \put(0,20){\line(1,0){10}}
  \put(0,50){\line(1,0){10}}
  \put(0,60){\line(1,0){10}}
  \put(0,70){\line(1,0){10}}

  \put(-30,5){\ldots}
  \put(-30,15){\ldots}
  \put(-30,25){\ldots}
  \put(-30,35){\ldots}
  \put(-30,45){\ldots}
  \put(-30,55){\ldots}
  \put(-30,65){\ldots}
\thicklines

  \put(-11.75,-3){\line(0,1){76}}
  \put(-8,-3){\line(0,1){76}}
  \put(-12.15,-3){\line(1,0){4.5}}
  \put(-12.15,73){\line(1,0){4.5}}

\end{picture}
  \caption{The two right-most columns of the dual lattice (using open
    boundary conditions) after the second step of reduction.}
  \label{fig:3}
\end{figure}

\begin{figure}[t]
  \centering
  \vspace{0.5cm}
  \begin{picture}(80,80)(-55,-15)

  \put(0,0){\circle*{3}}
  \put(10,0){\circle*{3}}
  \put(10,10){\circle*{3}}
  \put(0,10){\circle*{3}}

  \put(10,20){\circle*{3}}
  \put(0,20){\circle*{3}}

  \put(-1.75,35){\ldots}
 \put(10,50){\circle*{3}}
  \put(0,50){\circle*{3}}
 \put(10,60){\circle*{3}}
  \put(0,60){\circle*{3}}
\put(10,70){\circle*{3}}
  \put(0,70){\circle*{3}}

  \put(-10,0){\circle*{3}}
  \put(-10,10){\circle*{3}}
  \put(-10,20){\circle*{3}}

 \put(-10,50){\circle*{3}}
 \put(-10,60){\circle*{3}}
 \put(-10,70){\circle*{3}}

  \put(0,0){\line(-1,0){10}}
  \put(0,10){\line(-1,0){10}}
  \put(0,20){\line(-1,0){10}}
  \put(0,50){\line(-1,0){10}}
  \put(0,60){\line(-1,0){10}}
  \put(0,70){\line(-1,0){10}}

  \put(20,0){\circle*{3}}
  \put(20,10){\circle*{3}}
  \put(20,20){\circle*{3}}

  \put(20,50){\circle*{3}}
  \put(20,60){\circle*{3}}
  \put(20,70){\circle*{3}}

  \put(20,0){\line(-1,0){10}}
  \put(20,10){\line(-1,0){10}}
  \put(20,20){\line(-1,0){10}}
  \put(20,50){\line(-1,0){10}}
  \put(20,60){\line(-1,0){10}}
  \put(20,70){\line(-1,0){10}}

  \put(-30,5){\ldots}
  \put(-30,15){\ldots}
  \put(-30,25){\ldots}
  \put(-30,35){\ldots}
  \put(-30,45){\ldots}
  \put(-30,55){\ldots}
  \put(-30,65){\ldots}

  \put(-45,0){\circle*{3}}
  \put(-35,0){\circle*{3}}
  \put(-35,10){\circle*{3}}
  \put(-45,10){\circle*{3}}

  \put(-35,20){\circle*{3}}
  \put(-45,20){\circle*{3}}

   \put(-56.75,35){\ldots}
   \put(-35,50){\circle*{3}}
   \put(-45,50){\circle*{3}}
   \put(-35,60){\circle*{3}}
   \put(-45,60){\circle*{3}}
   \put(-35,70){\circle*{3}}
   \put(-45,70){\circle*{3}}

  \put(-55,0){\circle*{3}}
  \put(-55,10){\circle*{3}}
  \put(-55,20){\circle*{3}}

 \put(-55,50){\circle*{3}}
 \put(-55,60){\circle*{3}}
 \put(-55,70){\circle*{3}}

   \put(-45,0){\line(1,0){10}}
   \put(-45,10){\line(1,0){10}}
   \put(-45,20){\line(1,0){10}}
   \put(-45,50){\line(1,0){10}}
   \put(-45,60){\line(1,0){10}}
   \put(-45,70){\line(1,0){10}}

 \put(-65,0){\circle*{3}}
  \put(-65,10){\circle*{3}}
  \put(-65,20){\circle*{3}}

 \put(-65,50){\circle*{3}}
 \put(-65,60){\circle*{3}}
 \put(-65,70){\circle*{3}}

 \put(-65,0){\line(1,0){10}}
   \put(-65,10){\line(1,0){10}}
   \put(-65,20){\line(1,0){10}}
   \put(-65,50){\line(1,0){10}}
   \put(-65,60){\line(1,0){10}}
   \put(-65,70){\line(1,0){10}}

\end{picture}
  \caption{Reduced configuration, after completing the reduction
    process.}
  \label{fig:4}
\end{figure}
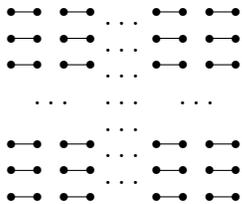

Consider a $2D$ $N \times N$ square lattice with open boundary
conditions.  In this case algorithm \ref{met1} is ergodic. The basic
idea of the proof is to perform transformations that ``shift'' all the
vertical active bonds to the left. This can be accomplished by
performing transformations according to the rules listed in
Table~\ref{tab:3}, starting from the upper-right corner of the dual
lattice, proceeding downward along a column of the dual lattice, and
then moving to the column to the left. We call this transformation
``reduction'', which we denote with ${\cal R}$. It is easily seen from
Table~\ref{tab:3} that it coincides with the identity if $A_2(x^*)=0$,
and with conjugation if $A_2(x^*)=1$, i.e.,
\begin{equation}
  \label{eq:s5}
 {\cal R}A(x^*)=\delta_{A_2,0}\,{\cal I}A(x^*) + \delta_{A_2,1}\,{\cal
   C} A(x^*).
\end{equation}
In the case of open boundary conditions, the sites of the dual lattice
are $x^*(i,j)$ with $i,j\in\{1,\ldots,N-1\}$. By construction, after
${\cal R}$ has been applied to the right-most column of the dual
lattice, the resulting configuration will not have any vertical bond
on the right-hand side of this column, i.e., $A_2(x^*)=0$ for
$x^*(i,N-1)$, $i\in\{1,\ldots,N-1\}$, and only horizontal bonds will
be present. As ${\cal R}$ transforms an admissible configuration into
another admissible configuration, the only possibility is that all the
horizontal bonds of the right-most column of the dual lattice are
active, as the rightmost sites of the direct lattice need to be
touched by at least one active bond and they cannot have more than one
(see Fig.~(\ref{fig:2})). We now apply ${\cal R}$ to the following
column: as the vertical bonds move to the left, and all the sites in
the before-last column of the direct lattice already have an active
bond, the horizontal bonds in the before-last column of the dual
lattice must be inactive (see Fig.~(\ref{fig:3})).  If we now repeat
the procedure, we find ourselves as after the first step: there is a
column of sites of the direct lattice with no vertical bonds, and no
horizontal bonds to the right, exactly as the right boundary of the
lattice. Therefore, the result iterates for pairs of columns of the
dual lattice, until we reach the left boundary. As all the sites are
already connected horizontally to the right, and as the uppermost site
can have at most a single vertical bond, it is easy to see that no
vertical bond can appear. We have then reduced the initial
configuration to the reduced configuration of Fig.~(\ref{fig:4}). As
no reference has been made to the specific form of the initial
configuration, the procedure applies equally to any admissible
configuration, which completes the proof of ergodicity for open
boundary conditions. This also provides an admissible configuration,
so completing the construction.

\begin{table}[h]
  \centering
  \begin{tabular}{c|c|c|c}
$S(x^*)$ &  $A(x^*)$  & $\hat{\cal R}S(x^*)$ & ${\cal R}A(x^*)$ \\ \hline
\begin{picture}(20,20)(0,-2.5)
  \put(0,0){\circle*{3}}
  \put(10,0){\circle*{3}}
  \put(10,10){\circle*{3}}
  \put(0,10){\circle*{3}}
\end{picture}
 & $(0,0,0,0)$ &
\begin{picture}(20,20)(0,-2.5)
  \put(0,0){\circle*{3}}
  \put(10,0){\circle*{3}}
  \put(10,10){\circle*{3}}
  \put(0,10){\circle*{3}}
\end{picture}
 & $(0,0,0,0)$
 \\ \hline
\begin{picture}(20,20)(0,-2.5)
  \put(0,0){\circle*{3}}
  \put(10,0){\circle*{3}}
  \put(10,10){\circle*{3}}
  \put(0,10){\circle*{3}}
  \put(0,0){\line(1,0){10}}
\end{picture}
  & $(1,0,0,0)$ &
\begin{picture}(20,20)(0,-2.5)
  \put(0,0){\circle*{3}}
  \put(10,0){\circle*{3}}
  \put(10,10){\circle*{3}}
  \put(0,10){\circle*{3}}
  \put(0,0){\line(1,0){10}}
\end{picture}
  & $(1,0,0,0)$ \\ \hline
\begin{picture}(20,20)(0,-2.5)
  \put(0,0){\circle*{3}}
  \put(10,0){\circle*{3}}
  \put(10,10){\circle*{3}}
  \put(0,10){\circle*{3}}
  \put(10,0){\line(0,1){10}}
\end{picture}
  & $(0,1,0,0)$ & 
\begin{picture}(20,20)(0,-2.5)
  \put(0,0){\circle*{3}}
  \put(10,0){\circle*{3}}
  \put(10,10){\circle*{3}}
  \put(0,10){\circle*{3}}
  \put(0,0){\line(1,0){10}}
  \put(10,10){\line(-1,0){10}}
  \put(0,10){\line(0,-1){10}}
\end{picture}
  & $(1,0,1,1)$
 \\ \hline
\begin{picture}(20,20)(0,-2.5)
  \put(0,0){\circle*{3}}
  \put(10,0){\circle*{3}}
  \put(10,10){\circle*{3}}
  \put(0,10){\circle*{3}}
  \put(10,10){\line(-1,0){10}}
\end{picture}
  & $(0,0,1,0)$ &
\begin{picture}(20,20)(0,-2.5)
  \put(0,0){\circle*{3}}
  \put(10,0){\circle*{3}}
  \put(10,10){\circle*{3}}
  \put(0,10){\circle*{3}}
  \put(10,10){\line(-1,0){10}}
\end{picture}
  & $(0,0,1,0)$
 \\ \hline
\begin{picture}(20,20)(0,-2.5)
  \put(0,0){\circle*{3}}
  \put(10,0){\circle*{3}}
  \put(10,10){\circle*{3}}
  \put(0,10){\circle*{3}}
  \put(0,10){\line(0,-1){10}}
\end{picture}
  & $(0,0,0,1)$ &
\begin{picture}(20,20)(0,-2.5)
  \put(0,0){\circle*{3}}
  \put(10,0){\circle*{3}}
  \put(10,10){\circle*{3}}
  \put(0,10){\circle*{3}}
  \put(0,10){\line(0,-1){10}}
\end{picture}
  & $(0,0,0,1)$ \\ \hline
\begin{picture}(20,20)(0,-2.5)
  \put(0,0){\circle*{3}}
  \put(10,0){\circle*{3}}
  \put(10,10){\circle*{3}}
  \put(0,10){\circle*{3}}
  \put(0,0){\line(1,0){10}}
  \put(10,0){\line(0,1){10}}
\end{picture}
  & $(1,1,0,0)$ &
\begin{picture}(20,20)(0,-2.5)
  \put(0,0){\circle*{3}}
  \put(10,0){\circle*{3}}
  \put(10,10){\circle*{3}}
  \put(0,10){\circle*{3}}
  \put(10,10){\line(-1,0){10}}
  \put(0,10){\line(0,-1){10}}
\end{picture}
  & $(0,0,1,1)$ \\ \hline
\begin{picture}(20,20)(0,-2.5)
  \put(0,0){\circle*{3}}
  \put(10,0){\circle*{3}}
  \put(10,10){\circle*{3}}
  \put(0,10){\circle*{3}}
  \put(0,0){\line(1,0){10}}
  \put(10,10){\line(-1,0){10}}
\end{picture}
  & $(1,0,1,0)$ &
\begin{picture}(20,20)(0,-2.5)
  \put(0,0){\circle*{3}}
  \put(10,0){\circle*{3}}
  \put(10,10){\circle*{3}}
  \put(0,10){\circle*{3}}
  \put(0,0){\line(1,0){10}}
  \put(10,10){\line(-1,0){10}}
\end{picture}
  & $(1,0,1,0)$ \\ \hline
\begin{picture}(20,20)(0,-2.5)
  \put(0,0){\circle*{3}}
  \put(10,0){\circle*{3}}
  \put(10,10){\circle*{3}}
  \put(0,10){\circle*{3}}
  \put(0,0){\line(1,0){10}}
  \put(0,10){\line(0,-1){10}}
\end{picture}
  & $(1,0,0,1)$ &
\begin{picture}(20,20)(0,-2.5)
  \put(0,0){\circle*{3}}
  \put(10,0){\circle*{3}}
  \put(10,10){\circle*{3}}
  \put(0,10){\circle*{3}}
  \put(0,0){\line(1,0){10}}
  \put(0,10){\line(0,-1){10}}
\end{picture}
  & $(1,0,0,1)$\\ \hline
\begin{picture}(20,20)(0,-2.5)
  \put(0,0){\circle*{3}}
  \put(10,0){\circle*{3}}
  \put(10,10){\circle*{3}}
  \put(0,10){\circle*{3}}
  \put(10,0){\line(0,1){10}}
  \put(10,10){\line(-1,0){10}}
\end{picture}
  & $(0,1,1,0)$ &
\begin{picture}(20,20)(0,-2.5)
  \put(0,0){\circle*{3}}
  \put(10,0){\circle*{3}}
  \put(10,10){\circle*{3}}
  \put(0,10){\circle*{3}}
  \put(0,0){\line(1,0){10}}
  \put(0,10){\line(0,-1){10}}
\end{picture}
  & $(1,0,0,1)$\\ \hline
\begin{picture}(20,20)(0,-2.5)
  \put(0,0){\circle*{3}}
  \put(10,0){\circle*{3}}
  \put(10,10){\circle*{3}}
  \put(0,10){\circle*{3}}
  \put(10,0){\line(0,1){10}}
  \put(0,10){\line(0,-1){10}}
\end{picture}
  & $(0,1,0,1)$ &
\begin{picture}(20,20)(0,-2.5)
  \put(0,0){\circle*{3}}
  \put(10,0){\circle*{3}}
  \put(10,10){\circle*{3}}
  \put(0,10){\circle*{3}}
  \put(10,0){\line(0,1){10}}
  \put(0,10){\line(0,-1){10}}
\end{picture}
  & $(0,1,0,1)$ \\ \hline
\begin{picture}(20,20)(0,-2.5)
  \put(0,0){\circle*{3}}
  \put(10,0){\circle*{3}}
  \put(10,10){\circle*{3}}
  \put(0,10){\circle*{3}}
  \put(10,10){\line(-1,0){10}}
  \put(0,10){\line(0,-1){10}}
\end{picture}
  & $(0,0,1,1)$ &
\begin{picture}(20,20)(0,-2.5)
  \put(0,0){\circle*{3}}
  \put(10,0){\circle*{3}}
  \put(10,10){\circle*{3}}
  \put(0,10){\circle*{3}}
  \put(10,10){\line(-1,0){10}}
  \put(0,10){\line(0,-1){10}}
\end{picture}
  & $(0,0,1,1)$ \\ \hline
\begin{picture}(20,20)(0,-2.5)
  \put(0,0){\circle*{3}}
  \put(10,0){\circle*{3}}
  \put(10,10){\circle*{3}}
  \put(0,10){\circle*{3}}
  \put(0,0){\line(1,0){10}}
  \put(10,0){\line(0,1){10}}
  \put(10,10){\line(-1,0){10}}
\end{picture}
  & $(1,1,1,0)$ &
\begin{picture}(20,20)(0,-2.5)
  \put(0,0){\circle*{3}}
  \put(10,0){\circle*{3}}
  \put(10,10){\circle*{3}}
  \put(0,10){\circle*{3}}
  \put(0,10){\line(0,-1){10}}
\end{picture}
  & $(0,0,0,1)$  \\ \hline
\begin{picture}(20,20)(0,-2.5)
  \put(0,0){\circle*{3}}
  \put(10,0){\circle*{3}}
  \put(10,10){\circle*{3}}
  \put(0,10){\circle*{3}}
  \put(0,0){\line(1,0){10}}
  \put(10,0){\line(0,1){10}}
  \put(0,10){\line(0,-1){10}}
\end{picture}
  & $(1,1,0,1)$ &
\begin{picture}(20,20)(0,-2.5)
  \put(0,0){\circle*{3}}
  \put(10,0){\circle*{3}}
  \put(10,10){\circle*{3}}
  \put(0,10){\circle*{3}}
  \put(10,10){\line(-1,0){10}}
\end{picture}
  & $(0,0,1,0)$ \\ \hline
\begin{picture}(20,20)(0,-2.5)
  \put(0,0){\circle*{3}}
  \put(10,0){\circle*{3}}
  \put(10,10){\circle*{3}}
  \put(0,10){\circle*{3}}
  \put(0,0){\line(1,0){10}}
  \put(10,10){\line(-1,0){10}}
  \put(0,10){\line(0,-1){10}}
\end{picture}
  & $(1,0,1,1)$ &
\begin{picture}(20,20)(0,-2.5)
  \put(0,0){\circle*{3}}
  \put(10,0){\circle*{3}}
  \put(10,10){\circle*{3}}
  \put(0,10){\circle*{3}}
  \put(0,0){\line(1,0){10}}
  \put(10,10){\line(-1,0){10}}
  \put(0,10){\line(0,-1){10}}
\end{picture}
  & $(1,0,1,1)$ \\ \hline
\begin{picture}(20,20)(0,-2.5)
  \put(0,0){\circle*{3}}
  \put(10,0){\circle*{3}}
  \put(10,10){\circle*{3}}
  \put(0,10){\circle*{3}}
  \put(10,0){\line(0,1){10}}
  \put(10,10){\line(-1,0){10}}
  \put(0,10){\line(0,-1){10}}
\end{picture}
  & $(0,1,1,1)$ &
\begin{picture}(20,20)(0,-2.5)
  \put(0,0){\circle*{3}}
  \put(10,0){\circle*{3}}
  \put(10,10){\circle*{3}}
  \put(0,10){\circle*{3}}
  \put(0,0){\line(1,0){10}}
\end{picture}
  & $(1,0,0,0)$\\ \hline
\begin{picture}(20,20)(0,-2.5)
  \put(0,0){\circle*{3}}
  \put(10,0){\circle*{3}}
  \put(10,10){\circle*{3}}
  \put(0,10){\circle*{3}}
  \put(0,0){\line(1,0){10}}
  \put(10,0){\line(0,1){10}}
  \put(10,10){\line(-1,0){10}}
  \put(0,10){\line(0,-1){10}}
\end{picture}
  & $(1,1,1,1)$ &
\begin{picture}(20,20)(0,-2.5)
  \put(0,0){\circle*{3}}
  \put(10,0){\circle*{3}}
  \put(10,10){\circle*{3}}
  \put(0,10){\circle*{3}}
\end{picture}
 & $(0,0,0,0)$\\ \hline
  \end{tabular}
\caption{Reduction: it coincides with the identity if $A_2=0$, and
  with conjugation if $A_2=1$, i.e., ${\cal R}=\delta_{A_2,0}\,{\cal
    I} + \delta_{A_2,1}\,{\cal C}$.}
\label{tab:3}
\end{table}

\subsection{Periodic boundary conditions for the $2D$ model}
\label{periodic2d}

\begin{figure}[t]
  \centering
  \begin{picture}(80,80)(-40,-15)
  \put(0,0){\circle*{3}}
  \put(10,0){\circle*{3}}
  \put(10,10){\circle*{3}}
  \put(0,10){\circle*{3}}

  \put(10,20){\circle*{3}}
  \put(0,20){\circle*{3}}

  \put(-18,76){\tiny $N\!\!-\!\!1$}
  \put(-2,76){\tiny $N$}
  \put(8.5,76){\tiny $1$}

  \put(-6.5,35){\ldots}
 \put(10,50){\circle*{3}}
  \put(0,50){\circle*{3}}
 \put(10,60){\circle*{3}}
  \put(0,60){\circle*{3}}
\put(10,70){\circle*{3}}
  \put(0,70){\circle*{3}}

  \put(-10,0){\circle*{3}}
  \put(-10,10){\circle*{3}}
  \put(-10,20){\circle*{3}}

 \put(-10,50){\circle*{3}}
 \put(-10,60){\circle*{3}}
 \put(-10,70){\circle*{3}}

  \put(0,0){\line(-1,0){10}}
  \put(0,10){\line(1,0){10}}
  \put(0,20){\line(1,0){10}}
  \put(0,50){\line(-1,0){10}}
  \put(0,60){\line(-1,0){10}}
  \put(0,70){\line(1,0){10}}

  \put(-30,5){\ldots}
  \put(-30,15){\ldots}
  \put(-30,25){\ldots}
  \put(-30,35){\ldots}
  \put(-30,45){\ldots}
  \put(-30,55){\ldots}
  \put(-30,65){\ldots}
\thicklines

  \put(-11.75,-3){\line(0,1){76}}
  \put(-8,-3){\line(0,1){76}}
  \put(-12.15,-3){\line(1,0){4.5}}
  \put(-12.15,73){\line(1,0){4.5}}

  \put(17,5){\ldots}
  \put(17,15){\ldots}
  \put(17,25){\ldots}
  \put(17,35){\ldots}
  \put(17,45){\ldots}
  \put(17,55){\ldots}
  \put(17,65){\ldots}

\put(12,-3){\line(0,1){76}}
  \put(8.25,-3){\line(0,1){76}}
  \put(12.35,-3){\line(-1,0){4.5}}
  \put(12.35,73){\line(-1,0){4.5}}

\end{picture}
  \caption{The $N$th and first columns of the dual lattice (using
    periodic boundary conditions) after the first step of
    reduction. (Numbers in the figure refer to columns of the direct
    lattice).}
  \label{fig:5}
\end{figure}
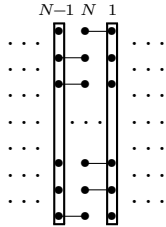

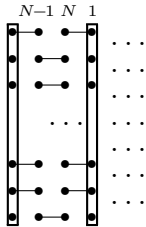
\begin{figure}[t]
  \centering
\vspace{0.5cm}
  \begin{picture}(80,80)(-40,-15)
  \put(0,0){\circle*{3}}
  \put(10,0){\circle*{3}}
  \put(10,10){\circle*{3}}
  \put(0,10){\circle*{3}}

  \put(10,20){\circle*{3}}
  \put(0,20){\circle*{3}}

  \put(-18,76){\tiny $N\!\!-\!\!1$}
  \put(-2,76){\tiny $N$}
  \put(8.5,76){\tiny $1$}

  \put(-6.5,35){\ldots}
 \put(10,50){\circle*{3}}
  \put(0,50){\circle*{3}}
 \put(10,60){\circle*{3}}
  \put(0,60){\circle*{3}}
\put(10,70){\circle*{3}}
  \put(0,70){\circle*{3}}

  \put(-10,0){\circle*{3}}
  \put(-10,10){\circle*{3}}
  \put(-10,20){\circle*{3}}

 \put(-10,50){\circle*{3}}
 \put(-10,60){\circle*{3}}
 \put(-10,70){\circle*{3}}

  \put(-20,0){\circle*{3}}
  \put(-20,10){\circle*{3}}
  \put(-20,20){\circle*{3}}

 \put(-20,50){\circle*{3}}
 \put(-20,60){\circle*{3}}
 \put(-20,70){\circle*{3}}

  \put(0,0){\line(-1,0){10}}
  \put(0,10){\line(1,0){10}}
  \put(0,20){\line(1,0){10}}
  \put(0,50){\line(-1,0){10}}
  \put(0,60){\line(-1,0){10}}
  \put(0,70){\line(1,0){10}}

  \put(-20,10){\line(1,0){10}}
  \put(-20,20){\line(1,0){10}}
  \put(-20,70){\line(1,0){10}}

\thicklines

  \put(-21.75,-3){\line(0,1){76}}
  \put(-18,-3){\line(0,1){76}}
  \put(-22.15,-3){\line(1,0){4.5}}
  \put(-22.15,73){\line(1,0){4.5}}

  \put(17,5){\ldots}
  \put(17,15){\ldots}
  \put(17,25){\ldots}
  \put(17,35){\ldots}
  \put(17,45){\ldots}
  \put(17,55){\ldots}
  \put(17,65){\ldots}

\put(12,-3){\line(0,1){76}}
  \put(8.25,-3){\line(0,1){76}}
  \put(12.35,-3){\line(-1,0){4.5}}
  \put(12.35,73){\line(-1,0){4.5}}

\end{picture}
  \caption{The $N-1$th, $N$th and first columns of the dual lattice
    (using periodic boundary conditions) after the first step of
    reduction.}
  \label{fig:6}
\end{figure}

Consider a $2D$ $N \times N$ square lattice with periodic boundary
conditions.  In this case, the sites of the dual lattice are
$x^*(i,j)$ with $i,j\in\{1,\ldots,N\}$, and are closed as well by
periodic boundary conditions. We start by applying ${\cal R}$ to
column $N$ of the dual lattice. Again, after this procedure there are
no vertical bonds active on the right-hand side of the column;
however, this time the right-most sites of the direct lattice can have
horizontal active bonds entering from the left or from the right, due
to the periodic boundary conditions (see Fig.~(\ref{fig:5})). There
are $2^N$ possibilities, according to the position of the horizontal
bonds (left or right). Repeating the procedure for column $N-1$, one
immediately sees that the active horizontal links must be the same as
in column 1, as there are no other possibilities due to the absence of
vertical bonds in column $N-1$ of the direct lattice (see
Fig.~(\ref{fig:6})).  Repeating again, we find for column $N-2$ the
same configuration of horizontal bonds of column $N$, and so on.
Since there is now an even number of columns on the dual lattice, the
final pattern will be $N/2$ identical pairs of columns on the dual
lattice, with all the odd columns equal to each other, and all the
even columns equal to each other.

\begin{figure}[t]
  \centering
\vspace{.5cm}
  \begin{picture}(80,80)(-40,-15)
  \put(0,0){\circle*{3}}
  \put(10,0){\circle*{3}}
  \put(10,10){\circle*{3}}
  \put(0,10){\circle*{3}}

  \put(10,20){\circle*{3}}
  \put(0,20){\circle*{3}}

  \put(-18,76){\tiny $N$}
  \put(-2,76){\tiny $1$}
  \put(8.5,76){\tiny $2$}

  \put(-6.5,35){\ldots}
 \put(10,50){\circle*{3}}
  \put(0,50){\circle*{3}}
 \put(10,60){\circle*{3}}
  \put(0,60){\circle*{3}}
\put(10,70){\circle*{3}}
  \put(0,70){\circle*{3}}

  \put(-10,0){\circle*{3}}
  \put(-10,10){\circle*{3}}
  \put(-10,20){\circle*{3}}

 \put(-10,50){\circle*{3}}
 \put(-10,60){\circle*{3}}
 \put(-10,70){\circle*{3}}

  \put(0,0){\line(-1,0){10}}
  \put(0,10){\line(1,0){10}}
  \put(0,20){\line(1,0){10}}
  \put(0,50){\line(-1,0){10}}
  \put(0,60){\line(-1,0){10}}
  \put(0,70){\line(1,0){10}}

  \put(-30,5){\ldots}
  \put(-30,15){\ldots}
  \put(-30,25){\ldots}
  \put(-30,35){\ldots}
  \put(-30,45){\ldots}
  \put(-30,55){\ldots}
  \put(-30,65){\ldots}
\thicklines

  \put(17,5){\ldots}
  \put(17,15){\ldots}
  \put(17,25){\ldots}
  \put(17,35){\ldots}
  \put(17,45){\ldots}
  \put(17,55){\ldots}
  \put(17,65){\ldots}

\end{picture}
  \caption{A possible almost-reduced configuration. The same pattern
    is repeated along the whole lattice.}
  \label{fig:7}
\end{figure}
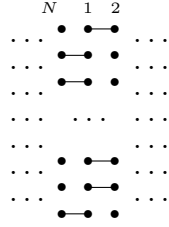

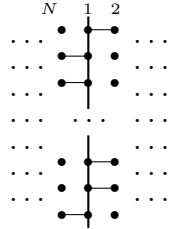
\begin{figure}[t]
  \centering
  \vspace{0.5cm}
  \begin{picture}(80,80)(-40,-15)
  \put(0,0){\circle*{3}}
  \put(10,0){\circle*{3}}
  \put(10,10){\circle*{3}}
  \put(0,10){\circle*{3}}

  \put(10,20){\circle*{3}}
  \put(0,20){\circle*{3}}

  \put(-18,76){\tiny $N$}
  \put(-2,76){\tiny $1$}
  \put(8.5,76){\tiny $2$}

  \put(-6.5,35){\ldots}
 \put(10,50){\circle*{3}}
  \put(0,50){\circle*{3}}
 \put(10,60){\circle*{3}}
  \put(0,60){\circle*{3}}
\put(10,70){\circle*{3}}
  \put(0,70){\circle*{3}}

  \put(-10,0){\circle*{3}}
  \put(-10,10){\circle*{3}}
  \put(-10,20){\circle*{3}}

 \put(-10,50){\circle*{3}}
 \put(-10,60){\circle*{3}}
 \put(-10,70){\circle*{3}}

  \put(0,0){\line(-1,0){10}}
  \put(0,10){\line(1,0){10}}
  \put(0,20){\line(1,0){10}}
  \put(0,50){\line(-1,0){10}}
  \put(0,60){\line(-1,0){10}}
  \put(0,70){\line(1,0){10}}

  \put(-30,5){\ldots}
  \put(-30,15){\ldots}
  \put(-30,25){\ldots}
  \put(-30,35){\ldots}
  \put(-30,45){\ldots}
  \put(-30,55){\ldots}
  \put(-30,65){\ldots}
\thicklines

 \put(0,-5){\line(0,1){35}}
\put(0,40){\line(0,1){35}}

  \put(17,5){\ldots}
  \put(17,15){\ldots}
  \put(17,25){\ldots}
  \put(17,35){\ldots}
  \put(17,45){\ldots}
  \put(17,55){\ldots}
  \put(17,65){\ldots}

\end{picture}
  \caption{Another possible almost-reduced configuration, with all the
    vertical bonds of column 1 in active state, winding around the
    lattice. Only the pattern of horizontal bonds is repeated along
    the whole lattice.}
  \label{fig:8}
\end{figure}

At first sight, one could think that there are therefore $2^N$ reduced
configurations instead of one (see Fig.~(\ref{fig:7})). At second
sight, one could think that there are even more, as it is possible
that all the vertical bonds in column 1 of the direct lattice are
active (see Fig.~(\ref{fig:8})): in this case all the sites in this
column have two more active bonds (vertical bonds are however absent
in the rest of the lattice, by construction).  However, one can show
that many of these configurations, that we call almost-reduced, can be
transformed into one another: by direct inspection, one can show that
a configuration with two horizontal bonds on the right of two adjacent
sites can be transformed into the configuration with two bonds on the
left of the same two adjacent sites, all the rest unchanged (and
vice-versa); and that a configuration with one bond on the right and
one on the left of two adjacent sites can be transformed into the
configuration with the first bond on the left and the second one on
the right of the same two adjacent sites, all the rest unchanged. In
the case without vertical bonds, it suffices to apply a conjugation to
a site of the dual lattice belonging to the row corresponding to the
two given sites, and then repeat the reduction procedure; in the case
with vertical bonds, it is enough to repeat the reduction procedure
along the whole lattice (see Fig.~(\ref{fig:8mezzo})).

Exploiting these equivalences, one can ``swap'' pairs of bonds,
finally reducing to one of the four configurations of
Fig.~(\ref{fig:9}). However, it is not possible to transform one of
these configurations into one another by means of our admissible
moves. To see this, define the number of active vertical bonds on row
$i$ of the dual lattice,
\begin{equation}
  \label{eq:parity}
  N_V(i)=\sum_{j=1}^{N} A_4(i,j) = \frac{1}{2}\sum_{j=1}^{N} A_4(i,j)
+  A_2(i,j)\,, \quad N_V(i)\in\mathbb{N}\,,
\end{equation}
and the number of active horizontal bonds on column $j$ of the dual lattice,
\begin{equation}
  \label{eq:parity2}
  N_H(j)=\sum_{i=1}^{N} A_1(i,j) = \frac{1}{2}\sum_{i=1}^{N} A_1(i,j)+
A_3(i,j)\,, \quad  N_H(j)\in\mathbb{N}\,.
\end{equation}
Observing that the admissible updates always change $A_1(i,j)+
A_3(i,j)$ and $A_4(i,j)+A_2(i,j)$ by $0$ or $\pm 2$, one has that the
{\it vertical parity} $P_V(i)= N_V(i)\,\mod 2$ and the {\it horizontal
  parity} $P_H(j)= N_H(j)\,\mod 2$ are conserved under admissible
moves. It is evident that for each of the four configurations of
Fig.~(\ref{fig:9}), $P_V(i)=P_V\,\forall i$ and of
$P_H(j)=P_H\,\forall j$, due to the horizontal periodicity of the
configuration, and moreover that these configurations have different
values of ($P_V,P_H$). As these quantities are conserved, we have that
$P_V(i)=P_V\,\forall i$ and $P_H(j)=P_H\,\forall j$ also for a generic
configuration, which under ${\cal R}$ will be transformed to the
reduced configuration with the same pair $(P_V,P_H)$.

\vspace{0.5cm}

\begin{figure}[htb]
  \centering
  \vspace{0.5cm}
  \begin{picture}(80,20)(-40,-15)
    \put(0,0){\circle*{3}}
    \put(0,10){\circle*{3}}

    \put(0,0){\line(1,0){10}}
    \put(0,10){\line(1,0){10}}

    \put(14,2.5){$\leftrightarrow$}

    \put(40,0){\circle*{3}}
    \put(40,10){\circle*{3}}

    \put(40,0){\line(-1,0){10}}
    \put(40,10){\line(-1,0){10}}

    \put(-10,0){\circle*{3}}
    \put(-10,10){\circle*{3}}
    \put(50,0){\circle*{3}}
    \put(50,10){\circle*{3}}

    \put(10,0){\circle*{3}}
    \put(10,10){\circle*{3}}

    \put(30,0){\circle*{3}}
    \put(30,10){\circle*{3}}

  \end{picture}
  \begin{picture}(80,20)(-40,-15)
    \put(0,0){\circle*{3}}
    \put(0,10){\circle*{3}}

    \put(0,0){\line(-1,0){10}}
    \put(0,10){\line(1,0){10}}

    \put(14,2.5){$\leftrightarrow$}

    \put(40,0){\circle*{3}}
    \put(40,10){\circle*{3}}

    \put(40,0){\line(1,0){10}}
    \put(40,10){\line(-1,0){10}}

    \put(-10,0){\circle*{3}}
    \put(-10,10){\circle*{3}}
    \put(50,0){\circle*{3}}
    \put(50,10){\circle*{3}}

    \put(10,0){\circle*{3}}
    \put(10,10){\circle*{3}}

    \put(30,0){\circle*{3}}
    \put(30,10){\circle*{3}}

  \end{picture}
  \begin{picture}(80,20)(-40,-15)
    \put(0,0){\circle*{3}}
    \put(0,10){\circle*{3}}

    \put(0,0){\line(1,0){10}}
    \put(0,10){\line(1,0){10}}

    \put(0,0){\line(0,1){10}}

    \put(14,2.5){$\leftrightarrow$}

    \put(40,0){\circle*{3}}
    \put(40,10){\circle*{3}}

    \put(40,0){\line(-1,0){10}}
    \put(40,10){\line(-1,0){10}}

    \put(40,0){\line(0,1){10}}

    \put(-10,0){\circle*{3}}
    \put(-10,10){\circle*{3}}
    \put(50,0){\circle*{3}}
    \put(50,10){\circle*{3}}

    \put(10,0){\circle*{3}}
    \put(10,10){\circle*{3}}

    \put(30,0){\circle*{3}}
    \put(30,10){\circle*{3}}

  \end{picture}
  \begin{picture}(80,20)(-40,-15)
    \put(0,0){\circle*{3}}
    \put(0,10){\circle*{3}}

    \put(0,0){\line(1,0){10}}
    \put(0,10){\line(-1,0){10}}

    \put(0,0){\line(0,1){10}}

    \put(14,2.5){$\leftrightarrow$}

    \put(40,0){\circle*{3}}
    \put(40,10){\circle*{3}}

    \put(40,0){\line(-1,0){10}}
    \put(40,10){\line(1,0){10}}

    \put(40,0){\line(0,1){10}}

    \put(-10,0){\circle*{3}}
    \put(-10,10){\circle*{3}}
    \put(50,0){\circle*{3}}
    \put(50,10){\circle*{3}}

    \put(10,0){\circle*{3}}
    \put(10,10){\circle*{3}}

    \put(30,0){\circle*{3}}
    \put(30,10){\circle*{3}}

  \end{picture}
  \caption{Allowed transformations of a pair of adjacent rows in the
    almost-reduced configurations. Only the basic block is drawn.}
  \label{fig:8mezzo}
\end{figure}
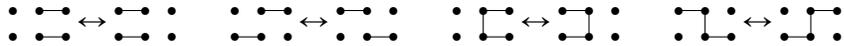

\begin{figure}[t]
  \centering
\vspace{0.5cm}
  \begin{picture}(80,80)(-40,-15)
  \put(0,0){\circle*{3}}
  \put(10,0){\circle*{3}}
  \put(10,10){\circle*{3}}
  \put(0,10){\circle*{3}}

  \put(10,20){\circle*{3}}
  \put(0,20){\circle*{3}}

  \put(-18,76){\tiny $N$}
  \put(-2,76){\tiny $1$}
  \put(8.5,76){\tiny $2$}

  \put(-14,94){$(0,0)$}

  \put(-6.5,35){\ldots}
 \put(10,50){\circle*{3}}
  \put(0,50){\circle*{3}}
 \put(10,60){\circle*{3}}
  \put(0,60){\circle*{3}}
\put(10,70){\circle*{3}}
  \put(0,70){\circle*{3}}

  \put(-10,0){\circle*{3}}
  \put(-10,10){\circle*{3}}
  \put(-10,20){\circle*{3}}

 \put(-10,50){\circle*{3}}
 \put(-10,60){\circle*{3}}
 \put(-10,70){\circle*{3}}

  \put(0,0){\line(1,0){10}}
  \put(0,10){\line(1,0){10}}
  \put(0,20){\line(1,0){10}}
  \put(0,50){\line(1,0){10}}
  \put(0,60){\line(1,0){10}}
  \put(0,70){\line(1,0){10}}

  \put(-30,5){\ldots}
  \put(-30,15){\ldots}
  \put(-30,25){\ldots}
  \put(-30,35){\ldots}
  \put(-30,45){\ldots}
  \put(-30,55){\ldots}
  \put(-30,65){\ldots}
\thicklines

  \put(17,5){\ldots}
  \put(17,15){\ldots}
  \put(17,25){\ldots}
  \put(17,35){\ldots}
  \put(17,45){\ldots}
  \put(17,55){\ldots}
  \put(17,65){\ldots}

\end{picture}
  \begin{picture}(80,80)(-40,-15)
  \put(0,0){\circle*{3}}
  \put(10,0){\circle*{3}}
  \put(10,10){\circle*{3}}
  \put(0,10){\circle*{3}}

  \put(10,20){\circle*{3}}
  \put(0,20){\circle*{3}}

  \put(-18,76){\tiny $N$}
  \put(-2,76){\tiny $1$}
  \put(8.5,76){\tiny $2$}

\put(-14,94){$(0,1)$}

  \put(-6.5,35){\ldots}
 \put(10,50){\circle*{3}}
  \put(0,50){\circle*{3}}
 \put(10,60){\circle*{3}}
  \put(0,60){\circle*{3}}
\put(10,70){\circle*{3}}
  \put(0,70){\circle*{3}}

  \put(-10,0){\circle*{3}}
  \put(-10,10){\circle*{3}}
  \put(-10,20){\circle*{3}}

 \put(-10,50){\circle*{3}}
 \put(-10,60){\circle*{3}}
 \put(-10,70){\circle*{3}}

  \put(0,0){\line(-1,0){10}}
  \put(0,10){\line(1,0){10}}
  \put(0,20){\line(1,0){10}}
  \put(0,50){\line(1,0){10}}
  \put(0,60){\line(1,0){10}}
  \put(0,70){\line(1,0){10}}

  \put(-30,5){\ldots}
  \put(-30,15){\ldots}
  \put(-30,25){\ldots}
  \put(-30,35){\ldots}
  \put(-30,45){\ldots}
  \put(-30,55){\ldots}
  \put(-30,65){\ldots}
\thicklines

  \put(17,5){\ldots}
  \put(17,15){\ldots}
  \put(17,25){\ldots}
  \put(17,35){\ldots}
  \put(17,45){\ldots}
  \put(17,55){\ldots}
  \put(17,65){\ldots}

\end{picture}
\begin{picture}(80,80)(-40,-15)
  \put(0,0){\circle*{3}}
  \put(10,0){\circle*{3}}
  \put(10,10){\circle*{3}}
  \put(0,10){\circle*{3}}

  \put(10,20){\circle*{3}}
  \put(0,20){\circle*{3}}

  \put(-18,76){\tiny $N$}
  \put(-2,76){\tiny $1$}
  \put(8.5,76){\tiny $2$}

\put(-14,94){$(1,0)$}

  \put(-6.5,35){\ldots}
 \put(10,50){\circle*{3}}
  \put(0,50){\circle*{3}}
 \put(10,60){\circle*{3}}
  \put(0,60){\circle*{3}}
\put(10,70){\circle*{3}}
  \put(0,70){\circle*{3}}

  \put(-10,0){\circle*{3}}
  \put(-10,10){\circle*{3}}
  \put(-10,20){\circle*{3}}

 \put(-10,50){\circle*{3}}
 \put(-10,60){\circle*{3}}
 \put(-10,70){\circle*{3}}

  \put(0,0){\line(1,0){10}}
  \put(0,10){\line(1,0){10}}
  \put(0,20){\line(1,0){10}}
  \put(0,50){\line(1,0){10}}
  \put(0,60){\line(1,0){10}}
  \put(0,70){\line(1,0){10}}

  \put(-30,5){\ldots}
  \put(-30,15){\ldots}
  \put(-30,25){\ldots}
  \put(-30,35){\ldots}
  \put(-30,45){\ldots}
  \put(-30,55){\ldots}
  \put(-30,65){\ldots}
\thicklines

 \put(0,-5){\line(0,1){35}}
\put(0,40){\line(0,1){35}}

  \put(17,5){\ldots}
  \put(17,15){\ldots}
  \put(17,25){\ldots}
  \put(17,35){\ldots}
  \put(17,45){\ldots}
  \put(17,55){\ldots}
  \put(17,65){\ldots}

\end{picture}
\begin{picture}(80,80)(-40,-15)
  \put(0,0){\circle*{3}}
  \put(10,0){\circle*{3}}
  \put(10,10){\circle*{3}}
  \put(0,10){\circle*{3}}

  \put(10,20){\circle*{3}}
  \put(0,20){\circle*{3}}

  \put(-18,76){\tiny $N$}
  \put(-2,76){\tiny $1$}
  \put(8.5,76){\tiny $2$}

\put(-14,94){$(1,1)$}

  \put(-6.5,35){\ldots}
 \put(10,50){\circle*{3}}
  \put(0,50){\circle*{3}}
 \put(10,60){\circle*{3}}
  \put(0,60){\circle*{3}}
\put(10,70){\circle*{3}}
  \put(0,70){\circle*{3}}

  \put(-10,0){\circle*{3}}
  \put(-10,10){\circle*{3}}
  \put(-10,20){\circle*{3}}

 \put(-10,50){\circle*{3}}
 \put(-10,60){\circle*{3}}
 \put(-10,70){\circle*{3}}

  \put(0,0){\line(-1,0){10}}
  \put(0,10){\line(1,0){10}}
  \put(0,20){\line(1,0){10}}
  \put(0,50){\line(1,0){10}}
  \put(0,60){\line(1,0){10}}
  \put(0,70){\line(1,0){10}}

  \put(-30,5){\ldots}
  \put(-30,15){\ldots}
  \put(-30,25){\ldots}
  \put(-30,35){\ldots}
  \put(-30,45){\ldots}
  \put(-30,55){\ldots}
  \put(-30,65){\ldots}
\thicklines

 \put(0,-5){\line(0,1){35}}
\put(0,40){\line(0,1){35}}

  \put(17,5){\ldots}
  \put(17,15){\ldots}
  \put(17,25){\ldots}
  \put(17,35){\ldots}
  \put(17,45){\ldots}
  \put(17,55){\ldots}
  \put(17,65){\ldots}

\end{picture}

  \caption{The four inequivalent reduced configurations for periodic
    boundary conditions. The pair $(P_V,P_H)$ is reported above each
    configuration.}
  \label{fig:9}
\end{figure}
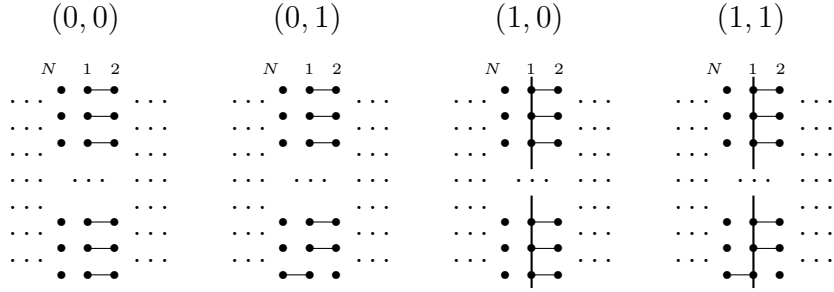

\subsection{Ergodicity in the $3D$ model}
\label{ergodic3d}

Consider a $3D$ cubic $N\times N\times N$ lattice, with $N$ even, and
a configuration of bonds satisfying the constraint that every site is
touched by an odd number of bonds. We define {\it plaquette update}
(PU) the process of inverting the state (active/inactive) of all the
bonds surrounding an elementary square of the lattice. We denote by
$P_{\mu\nu}(n)$ the square formed by the links $(n,n+\hat \mu)$,
$(n+\hat \mu, n+\hat\mu +\hat \nu)$, $(n+\hat\mu +\hat \nu,n+\hat
\nu)$ and $(n+\hat \nu,n)$, with $\mu,\nu=1,2,3$ and $n=(n_1,n_2,n_3)$
where $n_i=1,2,\ldots,N$.

We now show that by means of PU's it is possible to transform any
admissible configuration into any other in the case of open boundary
conditions (obc), while in the case of periodic boundary conditions
(pbc) it is necessary to supplement these transformations with a few
global transformations. In this way we can construct an ergodic
algorithm. The strategy is to reduce any admissible configuration to
one and the same, ``elementary'' configuration.

The proof is as follows. Consider the squares lying in the $(1,2)$
planes of the lattice. Fix $n_1=N-1$, and $\forall~n_2,n_3$ apply a PU
on the plaquette $P_{12}(n)$ if the rightmost bond (i.e., the one on
the link $(n+\hat 1, n+\hat 1 +\hat 2)$) is active, otherwise leave it
untouched. For obc $n_2=1,2,\ldots, N-1$ and $n_3=1,2,\ldots,N$, while
for pbc $n_2=1,2,\ldots, N$ and $n_3=1,2,\ldots,N$. After this
sequence of transformations, all the bonds on the links $(n+\hat 1,
n+\hat 1 +\hat 2)$ at $n_1=N-1$ are made inactive. Repeat now this
procedure for $n_1=N-2,N-3,\ldots,1$. As a result, all the bonds on
the links $(n+\hat 1, n+\hat 1 +\hat 2)$ (i.e. along direction $2$)
for $n_1=1,\ldots,N-1$, $n_2=1,\ldots, N-1$ (obc) or $n_2=1,\ldots, N$
(pbc) and $n_3=1,\ldots,N$ are made inactive.

Consider now the squares lying in the $(3,1)$ plane, and starting from
$n_1=N-1$, apply a PU on the plaquette $P_{31}(n)$ if the rightmost
bond (i.e., the one on the link $(n+\hat 1, n+\hat 1 +\hat 3)$) is
active, otherwise leave it untouched. For obc $n_3=1,2,\ldots, N-1$
and $n_2=1,2,\ldots,N$, while for pbc $n_3=1,2,\ldots, N$ and
$n_2=1,2,\ldots,N$; repeat then the procedure for
$n_1=N-2,N-3,\ldots,1$, so that in the end all the bonds on the links
$(n+\hat 1, n+\hat 1 +\hat 3)$ (i.e. along direction $3$) for
$n_1=1,\ldots,N-1$, $n_3=1,\ldots, N-1$ (obc) or $n_3=1,\ldots, N$
(pbc) and $n_2=1,\ldots,N$ are made inactive. This second series of
transformations clearly does not touch the bonds along direction $2$.

As a result, we have now a configuration where there are no active
bonds in the directions $2$ and $3$, except possibly at
$n_1=1$. Therefore, in the bulk of the lattice the constraint on the
number of bonds touching a site has to be enforced by means of bonds
in direction $1$, and therefore only one bond can touch a site. This
immediately implies the following relation:
\begin{equation}
  \label{eq:erg1}
  B_1(n_1,n_2,n_3) = 1-B_1(n_1+1,n_2,n_3)\,, \quad n_1=1,\ldots,N-2\,,
\end{equation}
for $n_2,n_3=1,\ldots,N$. For obc, since $B_1(N-1,n_2,n_3)=1$
$\forall~n_2,n_3$ in order for the sites at $n_1=N$ to satisfy the
constraint, this implies
\begin{equation}
  \label{eq:erg2}
{\rm obc:}\quad  B_1(n_1,n_2,n_3) = {\mod}(n_1,2)\,,\quad
n_1=1,\ldots,N-1\,,~\forall~n_2,n_3\,. 
\end{equation}
Since all sites at $n_1=1$ are touched by at least one bond, active
bonds in directions $2$ and $3$ at $n_1=1$ must form closed paths, so
that they contribute an even number of active bonds to all sites. Any
non-self-intersecting path of active bonds can be made inactive by
performing a PU on all the plaquettes contained in the path;
self-intersecting paths of active bonds can be decomposed in
non-self-intersecting paths that have no link in common, and so also
in this case all bonds in direction $2$ and $3$ can be made inactive,
i.e., we obtain $ B_{2,3}(n_1,n_2,n_3) = 0$ $\forall~n_1,n_2,n_3$.
This is the sought-after ``elementary configuration'', to which all
other configurations can be reduced in the case of obc.

For pbc, in order for the sites at $n_1=N$ to satisfy the constraint,
one has furthermore that $B_1(N-1,n_2,n_3) = 1-B_1(N,n_2,n_3)$,
$\forall~n_2,n_3$, and so $B_1(1,n_2,n_3)=1-B_1(N,n_2,n_3)$, which
implies that a single bond in direction $1$ touches the sites at
$n_1=1$. This again implies that the active bonds in directions $2$
and $3$ at $n_1=1$ must form closed paths. While for closed paths that
do not wind around the lattice the considerations made above apply, so
that the corresponding bonds can be made inactive, this is not true
for winding paths. However, we can basically repeat the same strategy
used above: starting from $n_2=N-1$, perform a PU in the plaquettes
$P_{23}(1,n_2,n_3)$ if the rightmost bond ($(n+\hat 2, n+\hat 2 +\hat
3)$) is active, for all $n_3$, and then repeat the procedure for all
$n_2$ until we reach $n_2=1$. At this point, there can be only bonds
along direction $2$, and possibly bonds along direction $3$ at
$n_2=1$. Enforcing the constraint that every site has to be touched by
an even number of bonds lying in the plane $n_1=1$, we conclude that
$B_2(1,n_2,n_3)=B_2(1,1,n_3)$ $\forall~n_2$, i.e., active bonds (if
any) in direction $2$ form closed straight-line paths winding around
the lattice.  As a consequence, the number of $2$-bonds touching a
site at $n_2=1$ is even (either zero or 2), so that the same must
apply for the $3$-bonds at $n_2=1$, and so again
$B_3(1,1,n_3)=B_3(1,1,1)$ $\forall~n_3$. Furthermore, it is easy to
see that the straight lines formed by the $2$-bonds can be parallelly
shifted, and that a pair of such straight lines at $n_3$ and $n_3+1$
can be made inactive, by means of PU's. In conclusion, it is always
possible to reduce the configuration of $2$- and $3$-bonds on the
plane $n_1=1$ to one of the following cases:
\begin{equation}
  \label{eq:erg3}
  \begin{aligned}
    1:&~ B_2(1,n_2,n_3)=0\,,\quad\forall~n_2,n_3\,,\\
    &~B_3(1,n_2,n_3)=0\,, \quad\forall~n_2,n_3\,;\\
    2:&~ B_2(1,n_2,1)=1\quad\forall~n_2\,, ~
    B_2(1,n_2,n_3)=0\quad\forall~n_2,\forall~n_3\ne 1\,,\\ 
    &~ B_3(1,n_2,n_3)=0\,, \quad\forall~n_2,n_3\,;\\
    3:&~ B_3(1,1,n_3)=1\quad\forall~n_3\,, ~
    B_3(1,n_2,n_3)=0\quad\forall~n_2\ne 1,\forall~n_3\,,\\ 
    &~ B_2(1,n_2,n_3)=0\,, \quad\forall~n_2,n_3\,;\\
    4:&~ B_2(1,n_2,1)=1\quad\forall~n_2\,, ~
    B_2(1,n_2,n_3)=0\quad\forall~n_2,\forall~n_3\ne 1\,,\\ 
    &~ B_3(1,1,n_3)=1\quad\forall~n_3\,, ~
    B_3(1,n_2,n_3)=0\quad\forall~n_2\ne 1,\forall~n_3\,.
  \end{aligned}
\end{equation}
The configurations $1-4$ are characterized by the number (0 or 1) of
$2$-bonds and $3$-bonds in a strip at fixed $n_2$ and $n_3$,
respectively. Notice that these numbers do not depend on which strip
we choose. Since there are no other $2$-bonds and $3$-bonds in the
rest of the lattice, these numbers are the same if we count the
$2$-bonds in a slice (i.e., all $n_1$ and $n_3$) at fixed $n_2$, and
if we count the $3$-bonds in a slice (i.e., all $n_1$ and $n_2$) at
fixed $n_3$. Again, these numbers do not depend on the chosen
slice. Since a PU in the $(\mu,\nu)$ plane does not change the parity
of the number of $\mu$- or $\nu$-bonds in a slice at fixed $n_\mu$ or
$n_\nu$ (although changing possibly their number), we can determine to
which of the above configurations in the $n_1=1$ plane a given generic
configuration can be reduced, by simply computing these parities,
which again do not depend on which slice we choose. Defining
\begin{equation}
  \label{eq:erg4}
  {\cal P}_2 = {\mod}(\sum_{n_1,n_3} B_2(n_1,n_2,n_3),2)\,, \quad
  {\cal P}_3 = {\mod}(\sum_{n_1,n_2} B_2(n_1,n_2,n_3),2)\,,
\end{equation}
one can easily classify the configurations $1-4$ (see
Table~\ref{tab:4}).
\begin{table}[t]
  \centering
  \begin{tabular}[h]{c|cc}
    conf. & ${\cal P}_2$ & ${\cal P}_3$\\
\hline
1 &  0  & 0   \\
2 &  1  & 0   \\
3 &  0  & 1   \\
4 &  1  & 1
  \end{tabular}
  \caption{Partial classification of reduced configurations in $3D$.}
  \label{tab:4}
\end{table}

The last step is to simplify as much as possible the configuration of
$1$-bonds. For each $(n_2,n_3)$, the configuration of $1$-bonds is
entirely determined by the value of $B_1(1,n_2,n_3)$. It is easy to
see that by means of PU's, we can simultaneously change the
configurations at $(n_2,n_3)$ and $(n_2+1,n_3)$ or $(n_2,n_3+1)$,
i.e., we can go from $B_1(1,n_2,n_3)=b_1$, $B_1(1,n_2+1,n_3)=b_2$
($b_i=0,1$) to $B_1(1,n_2,n_3)=1-b_1$, $B_1(1,n_2+1,n_3)=1-b_2$, or
from $B_1(1,n_2,n_3)=b_1$, $B_1(1,n_2,n_3+1)=b_2$ to
$B_1(1,n_2,n_3)=1-b_1$, $B_1(1,n_2,n_3+1)=1-b_2$. Using this
observation, we can change the position of those rows characterized by
$B_1(1,n_2,n_3)=0$, and trade a pair of such rows for a pair with
$B_1(1,n_2,n_3)=1$. For definiteness, we move them first towards
$n_2=1$ at fixed $n_3$, removing them when two show up at neighboring
sites: as a result, $B_1(1,n_2,n_3)=1$ $\forall~n_2\ne 1,n_3$. Next,
we move them towards $n_3=1$ at fixed $n_2=1$, again removing them
when two show up at neighboring sites: as a result, $B_1(1,n_2,n_3)=1$
$\forall~n_2\ne 1,n_3\ne 1$. Then, two possibilities remain: either
$B_1(1,1,1)=1$ or $B_1(1,1,1)=0$. This means that in the first case
there is an even number of $1$-bonds in any slice at fixed $n_1$,
while in the second case this number is odd. Since, as mentioned
above, the parity of the number of $1$-bonds in a slice at fixed $n_1$
is not changed by a PU, the ``reduced configuration'' of $1$-bonds
corresponding to any given generic configuration is determined by the
value of the quantity
\begin{equation}
  \label{eq:erg5}
    {\cal P}_1 = {\mod}(\sum_{n_2,n_3} B_1(n_1,n_2,n_3),2)\,.
\end{equation}
Therefore, the whole configuration space is made of 8 sectors, not
connected by PU's. Each sector is characterized by the values of the
parities ${\cal P}_i$, or, equivalently, by the ``reduced
configuration'' to which it can be brought by means of PU's alone. In
turn, these reduced configurations are entirely determined by the
following relations,
\begin{equation}
  \label{eq:erg6}
  \begin{aligned}
   {\rm pbc}:&~   B_1(n_1,n_2,n_3) = 1-B_1(n_1+1,n_2,n_3)\,, \quad
      n_1=1,\ldots,N-1\,,\\
      &~B_2(n_1,n_2,n_3)=B_3(n_1,n_2,n_3)=0\quad\forall~n_1\ne
      1,\forall~n_2\,,\forall~n_3\,,\\
      &~B_2(1,n_2,n_3)=0\,, \quad\forall~n_2\,,\forall~n_3\ne
      1\,,\\
      &~B_3(1,n_2,n_3)=0\,,\quad\forall~n_2\ne 1\,,\forall~n_3\,,\\
      &~B_2(1,n_2,1)=B_2(1,1,1)\,, \quad\forall~n_2\,,\\
      &~B_3(1,1,n_3)=B_3(1,1,1)\,,\quad\forall~n_3\,,
  \end{aligned}
\end{equation}
and by the values of $B_1(1,1,1)$, $B_2(1,1,1)$, and $B_3(1,1,1)$ (see
Table~\ref{tab:5}). In order to move from a sector to another, i.e.,
to change one of the parities ${\cal P}_i$ while remaining in the
space of admissible configurations, it is necessary to change the
state of all the bonds along a closed path winding around the lattice
along direction $i$.
\begin{table}[p]
  \centering
  \begin{tabular}[h]{c|ccc|ccc}
 &   $B_1(1,1,1)$ & $B_2(1,1,1)$ & $B_3(1,1,1)$ & ${\cal P}_1$ & ${\cal
      P}_2$ & ${\cal P}_3$\\ \hline
  \begin{picture}(30,30)(-15,-5)
  \put(0,0){\circle*{3}}
  \put(0,14){\circle*{3}}
  \put(10,8){\circle*{3}}
  \put(10,-4){\circle*{3}}
  \thicklines
  \put(0,0){\line(5,-2){10}}
\end{picture}& 1 & 0 & 0 & 0 & 0 & 0\\ \hline
    \begin{picture}(30,30)(-15,-5)
  \put(0,0){\circle*{3}}
  \put(0,14){\circle*{3}}
  \put(10,8){\circle*{3}}
  \put(10,-4){\circle*{3}}

  \thicklines
  
  \put(0,0){\line(5,-2){10}}
  \put(0,0){\line(0,1){14}}
\end{picture}
  &  1 & 0 & 1 & 0 & 0 & 1\\ \hline
\begin{picture}(30,30)(-15,-5)
  \put(0,0){\circle*{3}}
  \put(0,14){\circle*{3}}
  \put(10,8){\circle*{3}}
  \put(10,-4){\circle*{3}}

  \thicklines
  
  \put(0,0){\line(5,-2){10}}
  \put(0,0){\line(4,3){10}}
\end{picture}
  &  1 & 1 & 0 & 0 & 1 & 0\\ \hline
    \begin{picture}(30,30)(-15,-5)
  \put(0,0){\circle*{3}}
  \put(0,14){\circle*{3}}
  \put(10,8){\circle*{3}}
  \put(10,-4){\circle*{3}}

  \thicklines
  
  \put(0,0){\line(5,-2){10}}
  \put(0,0){\line(0,1){14}}
  \put(0,0){\line(4,3){10}}
\end{picture}
  &  1 & 1 & 1 & 0 & 1 & 1\\ \hline
    \begin{picture}(30,30)(-15,-5)
  \put(0,0){\circle*{3}}
  \put(0,14){\circle*{3}}
  \put(10,8){\circle*{3}}
  \put(10,-4){\circle*{3}}

  \thicklines
  
\end{picture}
  &  0 & 0 & 0 & 1 & 0 & 0\\ \hline
  \begin{picture}(30,30)(-15,-5)
  \put(0,0){\circle*{3}}
  \put(0,14){\circle*{3}}
  \put(10,8){\circle*{3}}
  \put(10,-4){\circle*{3}}

  \thicklines
  
  \put(0,0){\line(0,1){14}}
\end{picture}  &  0 & 0 & 1 & 1 & 0 & 1\\ \hline
\begin{picture}(30,30)(-15,-5)
  \put(0,0){\circle*{3}}
  \put(0,14){\circle*{3}}
  \put(10,8){\circle*{3}}
  \put(10,-4){\circle*{3}}

  \thicklines
  
\put(0,0){\line(4,3){10}}
\end{picture}  &  0 & 1 & 0 & 1 & 1 & 0\\ \hline
\begin{picture}(30,30)(-15,-5)
  \put(0,0){\circle*{3}}
  \put(0,14){\circle*{3}}
  \put(10,8){\circle*{3}}
  \put(10,-4){\circle*{3}}

  \thicklines
  
  \put(0,0){\line(0,1){14}}
  \put(0,0){\line(4,3){10}}
  \put(0,0){\line(4,3){10}}
\end{picture}  &  0 & 1 & 1 & 1 & 1 & 1 \\ \hline
  \end{tabular}
  \caption{Complete classification of the reduced configurations in
    $3D$.}
  \label{tab:5}
\end{table}

\section{Determination of the upper and 
lower number of bonds permitted in a given configuration (pbc case)}
\label{extremebonds}

Due to the constraints, the number of dimers touching a given site
must be odd for an admissible configuration.  In dimension $D$ and
using periodic boundary conditions, this means that this number is
$1,3 \ldots, 2D - 1$.  We denote by $V_{k}$ the number of vertices
with $k$ dimers in a given configuration, and by $V$ the total number
of sites. One has the two following relations:
\begin{equation}
\begin{aligned}
\sum_{k=1}^{D} V_{2k-1}&=V, \\
\sum_{k=1}^{D}(2k-1)V_{2k-1}&= 2B,
\end{aligned}
\end{equation} 
where $B$ is the total number of dimers. 
The first equation above can be rewritten as 
\begin{equation}
V=V_{1} + \sum_{k=2}^{D}V_{2k-1}
\end{equation}
and substituting this into the second equation we get 
\begin{equation}
2B=V+\sum_{k=2}^{D} 2(k-1)V_{2k-1} \leq V + \sum_{k=1}^{D}2(D-1)V_{2k-1}.
\label{2B}
\end{equation}
Since the terms under the summation signs are positive, we get the inequalities 
\begin{equation}
V\leq 2B \leq (2D-1)V
\end{equation}
and dividing by the total number of links $DV$ we obtain
\begin{equation}
\frac{1}{2D}\leq \frac{B}{DV}\leq 1 - \frac{1}{2D}\,.
\end{equation}

\section{Average number of bonds (pbc)}
\label{averagebonds}

Given an admissible configuration of dimers $b=\{ A_i(x^*) \}$ one
immediately sees that the configuration $A_i^{'}(x^*) = 1 - A_i(x^*)$
is still admissible. Indeed, the change in the number of dimers
$\pi_x[b]$ touching site $x$ is $\pi_x[b']-\pi_x[b]=2(D-\pi_x[b])$,
which is even, so that if $\pi_x[b]$ is odd, then $\pi_x[b']$ is
odd. Setting $t=\tanh F$ we can therefore write for the partition
function (up to an irrelevant factor)
\begin{equation}
Z=\sum_{\{ b \in \tilde{\cal B}\}}t^{B}=\sum_{\{ b \in \tilde{\cal B}\}}t^{DV-B}=
\sum_{\{ b \in \tilde{\cal B}\}}t^{B}t^{DV-2B}
\end{equation}
or equivalently 
\begin{equation}
\lla t^{DV-2B}\rra=1\,.
\end{equation}
Using the well-known inequality $\langle e^{A} \rangle \geq e^{\langle
  A \rangle}$, we find
\begin{equation}
t^{DV - 2 \lla B \rra} \leq 1\,,
\end{equation}
which, since $t \leq 1$, implies 
\begin{equation}
DV - 2 \lla B \rra \geq 0 \Rightarrow \lla
\frac{B}{DV} \rra \leq \frac{1}{2}.
\end{equation}

\section{Analytic evaluation of the spin-spin correlation functions}
\label{analytic}
  
Here we show the analytic results for the $2D$ ferromagnetic Ising
model with imaginary magnetic field $\frac{H}{kT} = i\frac{\pi}{2}$
that have been derived in Ref.~\cite{Wu}.  The most important result
for our purposes is the asymptotic behavior of the spin-spin
correlation function. Since in the geometric representation the weight
of each graph depends only on $t = \tanh |F|$, the partition function
is the same in the ferromagnetic and antiferromagnetic cases;
furthermore, independently of the sign of the coupling, the spin-spin
correlation functions read
\begin{equation}
\langle s_{0}s_{d} \rangle = \langle \langle {\tanh F}^{d-2N_{d}} \rangle \rangle
\end{equation}
where $N_{d}$ is the number of active bonds on the straight-line path connecting the sites 
$(0,0)$ and $(d,0)$. Therefore we can write
\begin{equation}
\langle s_{0}s_{d} \rangle = (\text{sgn} F)^{d} \langle \langle t^{d-2N_{d}} \rangle \rangle = 
\begin{cases}
 \phantom{(-1)^{d}} \langle \langle t^{d-2N_{d}} \rangle \rangle & F > 0\,,  \\
 (-1)^{d}  \langle \langle t^{d-2N_{d}} \rangle \rangle & F < 0\,,
\end{cases}
\end{equation}
i.e., $\langle s_{0}s_{d} \rangle_{AFM} = (-1)^{d}\langle s_{0}s_{d}
\rangle_{FM}$.
The result obtained in Ref.~\cite{Wu} for $\langle s_{0}s_{d}
\rangle_{FM}$ is the following:
\begin{equation}
\langle s_{0}s_{d} \rangle_{FM} = M^{2} \left \{ 1 -
\frac{(-1)^{d}}{4\pi}\frac{(1-t^{2})^{2}}{t(1+t^{2})}\frac{1}{d} \left
( \frac{1-t}{1+t}\right )^{2d} \right \}\,, \nonumber
\end{equation}

\begin{equation}
M = \frac{1}{2^{\frac{3}{8}}}\frac{(1+t^{2})^{\frac{1}{2}}}{t^{\frac{1}{4}}(1+t^{4})^{\frac{1}{8}}}\,,
\label{staggered_magn}
\end{equation}
where $M$ is the staggered magnetization.
Then the result for the antiferromagnetic coupling immediately follows: 
\begin{equation}
\langle s_{0}s_{d} \rangle_{AFM} = M^{2} \left \{ (-1)^{d} - \frac{1}{4\pi}\frac{(1-t^{2})^{2}}{t(1+t^{2})}\frac{1}{d} \left ( \frac{1-t}{1+t}\right )^{2d} \right \}\,.
\end{equation}

\end{document}